\documentclass[prl,twocolumn,superscriptaddress,nobibnotes,a4paper]{revtex4-2}

\usepackage[pdftex]{graphicx}
\usepackage[pdftex]{epsfig}
\usepackage{amsmath}
\usepackage{amssymb}
\usepackage{amsfonts}
\usepackage{color}
\usepackage{hhline}
\usepackage{tabularx}
\usepackage{upgreek}
\usepackage{soul}

\usepackage{array} 
\usepackage[colorlinks=true, allcolors=blue]{hyperref} 

\newcommand{\Mech}{\mathrm{M}}
\newcommand{\OM}{\mathrm{OM}}
\newcommand{\cav}{\mathrm{cav}}
\newcommand{\PhC}{\mathrm{PhC}}

\begin{document}

\title{Integrated optical-readout of a high-Q mechanical out-of-plane mode}

\author{Jingkun Guo and Simon Gr\"oblacher}
\email{s.groeblacher@tudelft.nl}
\affiliation{Kavli Institute of Nanoscience, Department of Quantum Nanoscience, Delft University of Technology, 2628CJ Delft, The Netherlands}

\begin{abstract}
    The rapid development of high-$Q_\Mech$ macroscopic mechanical resonators has enabled great advances in optomechanics. Further improvements could allow for quantum-limited or quantum-enhanced applications at ambient temperature. Some of the remaining challenges include the integration of high-$Q_\Mech$ structures on a chip, while simultaneously achieving large coupling strengths through an optical read-out. Here, we present a versatile fabrication method, which allows us to build fully integrated optomechanical structures. We place a photonic crystal cavity directly above a mechanical resonator with high-$Q_\Mech$ fundamental out-of-plane mode, separated by a small gap. The highly confined optical field has a large overlap with the mechanical mode, enabling strong optomechanical interaction strengths. Furthermore, we implement a novel photonic crystal design, which allows for a very large cavity photon number, a highly important feature for optomechanical experiments and sensor applications. Our versatile approach is not limited to our particular design but allows for integrating an out-of-plane optical read-out into almost any device layout. Additionally, it can be scaled to large arrays and paves the way to realizing quantum experiments and applications with mechanical resonators based on high-$Q_\Mech$ out-of-plane modes alike.
\end{abstract}

\maketitle

Integrated cavity optomechanical systems have attracted significant attention for their potential use in both classical~\cite{Krause2012,Allain2020,Westerveld2021} and quantum~\cite{Stannigel2010,Wallucks2020,Fiaschi2021} applications, and for their ability to study fundamental physics~\cite{Bahrami2014,Carlesso2019}. A mechanical resonator with high quality factor ($Q_\Mech$) and large coupling to the optical cavity field is desirable, as they are directly related to the ability to maintain coherence and to have efficient readout and control~\cite{Aspelmeyer2014}. Additionally, in order to enable practical applications and advanced quantum experiments, an optomechanical system fully integrated on a chip is required.

In recent years, significant progress has been made in designing and fabricating high-$Q_\Mech$ integrated mechanical resonators. In particular, out-of-plane mechanical modes with ultra high quality factors have been demonstrated~\cite{Norte2016,Tsaturyan2017,Ghadimi2018,Hoej2021,Beccari2021}. Recently, the regime $Q_\Mech \cdot f_\Mech > k_B T / h$ has been achieved even at room temperature, where $f_\Mech$ is the resonance frequency of the mechanical mode, $T$ is the bath temperature, and $k_B$ and $h$ are the Boltzmann and the Planck constant. Within this regime, the thermal decoherence time is longer than their oscillation periods~\cite{Aspelmeyer2014}. This allows to perform quantum limited sensing~\cite{Norte2018,Whittle2021} or observe macroscopic quantum phenomena~\cite{Gut2020} at high temperature, if the mechanical resonator can also be measured in an efficient way~\cite{Wilson2015}. The quantum cooperativity $C_\mathrm{qu}$, which compares the measurement rate to the thermal decoherence rate, and should be around or larger than 1, gives a direct benchmark for reaching the regime of efficient readout and the potential to perform quantum experiments.

\begin{figure*}[th]
	\centering
	\includegraphics[width=1.\textwidth]{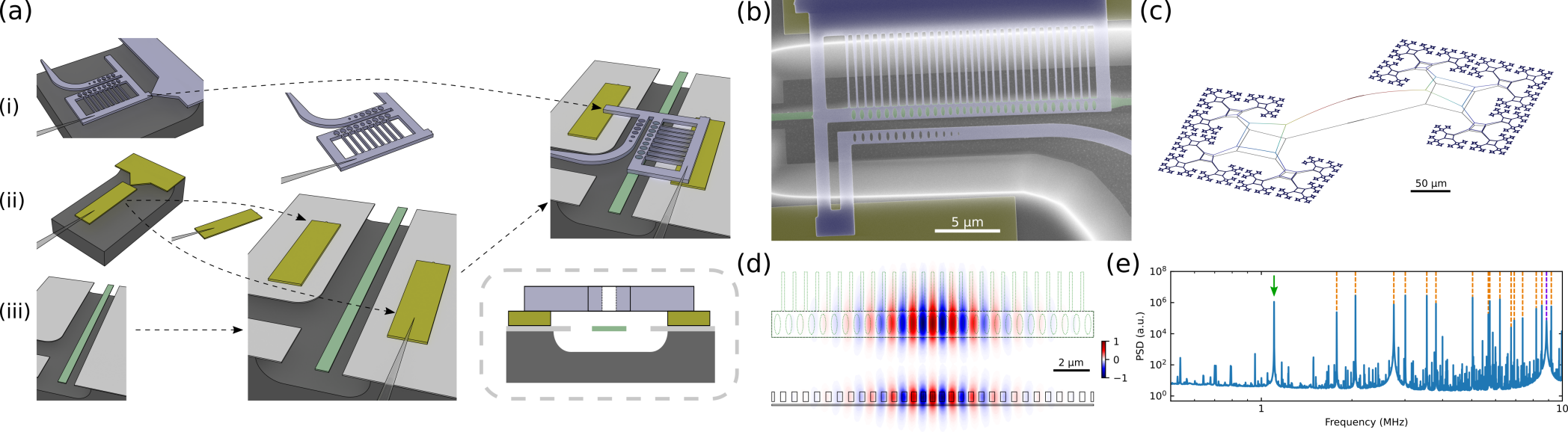}
	\caption{(a) Device assembly process. We start with 3 chips, on which the different structures are fabricated from thin film silicon nitride. They are patterned with (i) a photonic crystal, (ii) spacers, and (iii) a mechanical structure, respectively. A fiber is used to first pick up the spacers and place them on the mechanical chip. Then the photonic crystal is stacked on top of the spacers. The sketch in the box at the bottom-right shows the cross-section of the fully assembled structure. The thickness of the spacers determines the gap distance between the photonic crystal and the mechanical resonator. (b) False-colored SEM image of the assembled device. (c) Simulation of the fundamental out-of-plane mechanical mode of the fractal structure. The mechanical motion is gradually damped from the center to the clamping points. (d) Electric field distribution (top) on the upper surface of the mechanical layer and (bottom) on the center vertical plane. The green dashed line on the upper plot shows the projection of the photonic crystal. The mechanical layer in the simulation has a width of 1~$\upmu$m, matching the width of the photonic crystal. (e) Measured power spectral density (PSD) of an assembled device, showing how clean the spectrum is around the fundamental mode, with negligible excess noise from other modes. The green arrow indicates the fundamental high $Q_\Mech$ mechanical mode, the orange dashed lines the higher mechanical modes, while the mechanical mode of the photonic crystal is marked in purple. All other peaks are electronic noise.}
	\label{fig:DevGeneral}
\end{figure*}

While large mechanical quality factors have been shown in many different systems, coupling these mechanical modes to an integrated optical cavity remains challenging, which limits their potential use in optomechanical applications. To date, there is a disconnect between the largest mechanical quality factors, which are usually out-of-plane modes, and largest $C_\mathrm{qu}$, which is either achieved in-plane or with bulk-optics setups. In general, out-of-plane motion has the potential to provide the highest $Q_\Mech \cdot f_\Mech$ due to the possibility of minimizing the material thickness in the direction of motion~\cite{Unterreithmeier2010,Ghadimi2018}, which reduces clamping loss, with even the fundamental mode exhibiting high quality factors~\cite{Fedorov2020,Beccari2021}. Furthermore, out-of-plane modes can achieve a larger surface area perpendicular to the motional direction, which allows for easier coupling to external systems. Both these features make them highly interesting for potential sensing applications~\cite{Albrecht1991,Carlesso2018,Pate2020,Haelg2021}. Mechanical structures with a high-$Q_\Mech$ fundamental out-of-plane mode could even help minimize the disturbance from higher order modes, providing a spectrally clean platform for further quantum optomechanical experiments~\cite{Muhonen2019,Guo2019,Gut2020,Galinskiy2020}. Several attempts on making integrated optomechanical devices that couple to the out-of-plane motion have been made~\cite{Wilson2015,Zobenica2017,Liu2020}. However, forming a fully integrated optomechanical device with high-$Q_\Mech$ mechanical resonator and a large optomechanical coupling remains an outstanding hurdle to practical applications and novel quantum experiments.

In this work, we develop a versatile and flexible new fabrication method enabling the integration of large optomechanical coupling to a high-$Q_\Mech$ out-of-plane mechanical mode. In particular, we demonstrate devices efficiently coupled to the high-$Q_\Mech$ fundamental mechanical mode. Our method is based on a pick-and-place technique~\cite{Elshaari2020,Marinkovic2021}, allowing us to fabricate structures where a photonic crystal (PhC) is placed above a mechanical resonator with a 1.1~MHz fundamental out-of-plane mode and an intrinsic quality factor of around $2 \times 10^7$ at room temperature. The resulting devices exhibit a clean mode spectrum around the mode of interest. The photonic crystal and the mechanical structure are separated by a controllable small gap of around 100~nm. For a spacing of 130~nm, it is possible to achieve an optomechanical coupling rate $g_0/2\pi \approx 260$~kHz and $Q_\Mech \approx 1.6 \times 10^7$ simultaneously, corresponding to $f_\Mech \cdot Q_\Mech \approx 2.9 ~ (k_\mathrm{B} T/ h)$ at room temperature. We further show that the structure allows to use a large intracavity photon number, which leads to a strongly light-enhanced optomechanical coupling $g$ and allows to approach unity quantum cooperativity $C_\mathrm{qu}$~\cite{Aspelmeyer2014,Wilson2015} at high temperature. Our novel technique provides a highly versatile platform for future quantum experiments and applications with high-$Q_\Mech$ out-of-plane mechanical motion.

The fabrication of our devices is based on a pick-and-place method~\cite{Elshaari2020,Marinkovic2021}. As shown in Figure~\ref{fig:DevGeneral}(a), we first individually fabricate the mechanical and photonic structures, as well as the spacers out of silicon nitride on three separate chips. The spacers are used to provide support to the photonic crystal and to define the gap size between the mechanics and the cavity. The spacers (typical size of 28 $\upmu$m $\times$ 6 $\upmu$m) and the photonic crystal attach to the original substrate via a weak tether, with a width of about 100~nm. A tapered optical fiber with a sharp tip is then placed on the spacers / PhC, which adheres to the fiber through van der Waals and electrostatic forces. By moving the fiber, it is now possible to break the weak tether, while the structure remains attached to the fiber. We first pick up the spacers and position them on the mechanical chip. Then, a photonic crystal structure is transferred and placed above the spacers, as the top layer. When performing the picking and placing, the target chip is on a stage with rotation and three position degrees of freedom. We monitor this process through a camera attached to a microscope. This simple optical imaging is sufficient to achieve good alignment, as can be seen from the optical images taken during the transferring process (Figure~\ref{fig:assembpics}) and the scanning electron microscope (SEM) image of the final device (Figure~\ref{fig:DevGeneral}(b)).

Our mechanical design is inspired by a fractal-like structure~\cite{Fedorov2020,Beccari2021} (see Figure~\ref{fig:DevGeneral}(c)), which has been proposed and shown to have an extraordinary high mechanical quality factor of the fundamental mode for low frequency mechanics. A central string is connected to a block on each side and each unit is then connected to three similar, but smaller, sub (or \textit{child}) units. Our mechanical structure is fabricated from 50~nm thick high stress silicon nitride, where the tensile stress contributes to the large mechanical quality factor~\cite{Southworth2009,Unterreithmeier2010,Ghadimi2018}. In our structure, an additional child unit facilitates preserving the high stress in the parent unit when compared to the binary tree in~\cite{Fedorov2020}. The stress at the center can be maintained or even enhanced without significantly increasing the width of the tethers in the child units. In our structures, the width of the tethers slightly increases from 500~nm at the center to 1.4~$\upmu$m at the clamping points. In simulations, this results in an increased stress of 1.6~GPa at the center tether, up from the 1.3~GPa intrinsic stress. The smaller tether width in our structures further simplifies the silicon undercut process with a fluorine-based dry release method~\cite{Norte2018,Guo2019,Groeblacher2021}. In our design, the parent and the child units are connected by four tethers forming a diamond shape. Each tether connects to only two other tethers. Moving from the center of our structure to the clamping region, the gradient of the mechanical motion displacement can be suppressed gradually at each branching point. The bending is then distributed over the branching regions, forming a softly clamped structure resulting in reduced losses~\cite{Fedorov2020}.

The optical read-out cavity is formed by a separate photonic crystal cavity, made from silicon nitride with a width of about 1~$\upmu$m. The structure and the simulation of the electric field are shown in Figure~\ref{fig:DevGeneral}(b, d). The photonic crystal cavity strongly confines the light at the center. Extra tethers on the side raise the frequency of its mechanical modes (Figure~\ref{fig:MechSpectMechModes}(a)), minimizing its impact onto the mechanical noise spectrum. The clamping tethers further improve the thermal conductivity and the rigidity of our photonic crystal, allowing us to use a large cavity photon number without entering the thermal bi-stability regime~\cite{Almeida2004,Aspelmeyer2014,Camacho2009}. The evanescent field couples to the mechanical structure, as shown in the bottom of Figure~\ref{fig:DevGeneral}(d). By minimizing the distance between the photonic crystal and the mechanical structure, a large optomechanical coupling can be obtained. To further increase the coupling, the part of mechanical structure below the photonic crystal is widened in order to maximize the overlap with the optical field. In particular, we choose a width of 0.8~$\upmu$m and 1.0~$\upmu$m, similar to the width of the photonic crystal.

A broad-band spectrum, measured after assembling the device, is also shown in Figure~\ref{fig:DevGeneral}(e), where the fundamental mechanical mode is marked with a green arrow. Higher order modes are far away from the fundamental mode, where the lowest mechanical peak appears at 1.8~MHz, and the suspended photonic crystal has a fundamental mechanical mode at 9~MHz. Such large relative frequency spacing is extremely challenging to achieve with standard structures with directly integrated phononic crystal cavities~\cite{Tsaturyan2017,Ghadimi2018,Guo2019}, while it is crucial for many potential experiments~\cite{Muhonen2019,Guo2019,Gut2020,Galinskiy2020}.

\begin{figure}[!t]
    \centering
    \includegraphics[width=1.\columnwidth]{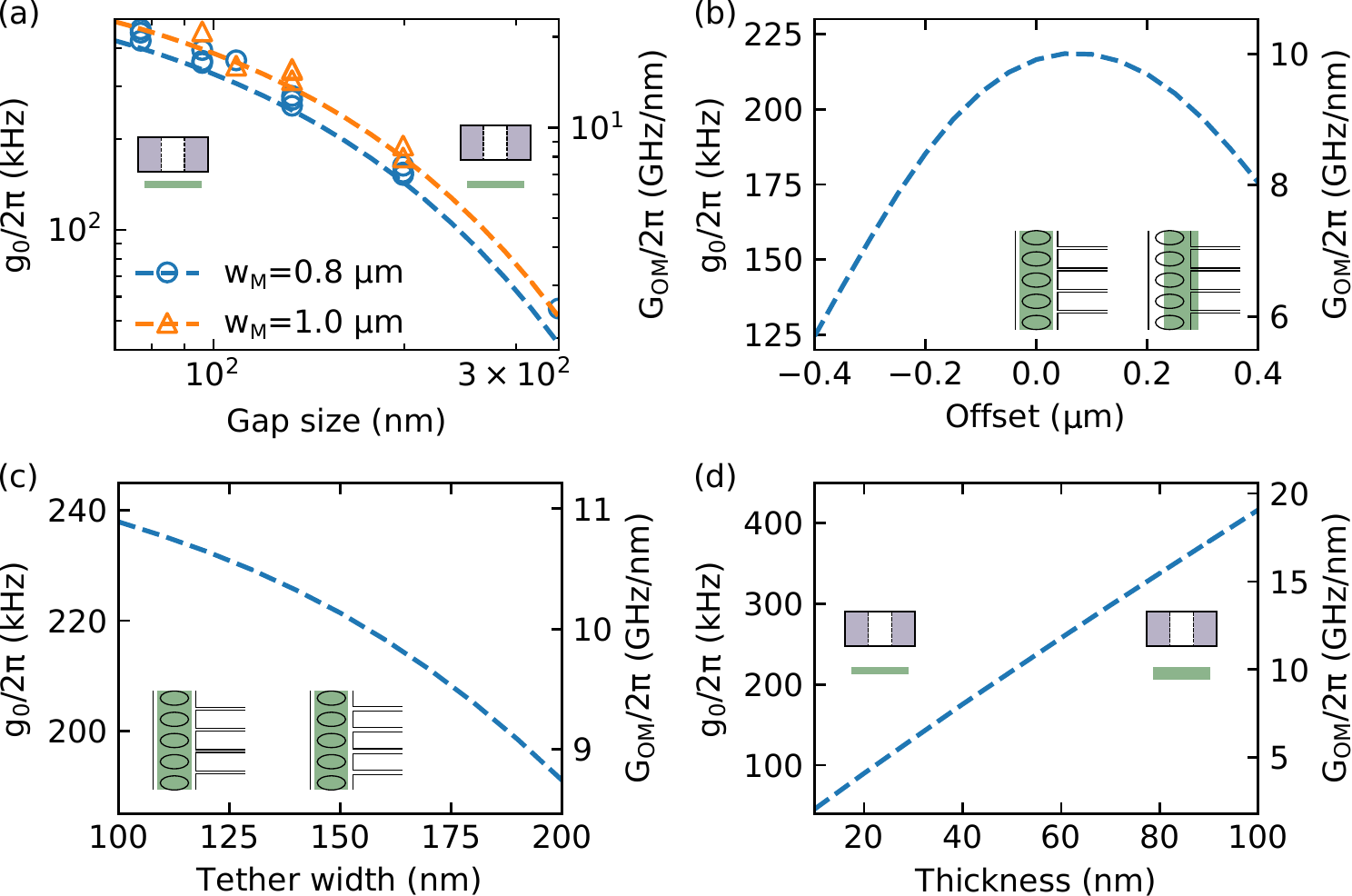}
    \caption{Optomechanical coupling rate $g_0$ and strength $G_\OM$ for different (a) gap sizes, (b) misalignment or offsets and (c) widths of the clamping tethers, and (d) thickness of the mechanical layer. The dashed lines are simulations, while the circles and triangles in (a) are measurement results for different mechanical layer thickness (blue and orange).}
    \label{fig:g0_Param}
\end{figure}

Achieving a small gap between the mechanical and photonic layers is important to achieve a high optomechanical coupling. This is especially true in our case as the mechanical resonator only couples to the evanescent field from the photonic crystal. We calculate the optomechanical coupling strength $G_\OM$ using the perturbation method~\cite{Johnson2002} and the corresponding optomechanical coupling rate $g_0$~\cite{Aspelmeyer2014} for our structure, as shown in Figure~\ref{fig:g0_Param}. Increasing the gap size from 75~nm to 350~nm reduces the optomechanical coupling strength $G_\OM/2\pi$ by one order of magnitude, from 21.6~GHz/nm to 2.2~GHz/nm, for a mechanical resonator width of 1~$\upmu$m. With our pick-and-place fabrication technique there is in principle no real lower limit on the gap size, as it is defined by the spacer. Furthermore, as it is performed in air and no further processing is required afterwards, there are no adhesion issues~\cite{Scheeper1992} or risks of collapsing structures. In our setup we reliably achieve a gap of 75~nm. We compare the measurements on our devices to the simulations of the optomechanical coupling, which shows good agreement. As simulated in Figure~\ref{fig:g0_Param}(b), the misalignment tolerance is also large and our alignment in the lateral direction is sufficiently good to achieve a large optomechanical coupling with this scheme. While a wider PhC structure would allow for an even larger tolerance, it would also lower the mechanical frequency of a low-$Q_\Mech$ torsional mode (cf.\ Figure~\ref{fig:MechSpectMechModes}(b)), which would introduce additional mechanical noise.

In this work, we use an asymmetric design of the photonic crystal, and hence a perfect alignment in the lateral direction (zero offset) does not give the best optomechanical coupling, as the clamping tethers attract the optical field to the side. Thus, a slight offset in our assembly improves the $g_0$ slightly. The value of the optomechanical coupling is also sensitive to the width of the clamping tethers. As shown in Figure~\ref{fig:g0_Param}(c), when there is no lateral offset, a reduction in the tether width increases the optomechanical coupling. Another factor for the optomechanical coupling is the thickness of the mechanical layer, shown in Figure~\ref{fig:g0_Param}(d). For our structure, $G_\OM$ depends on the electric field difference between the upper and the lower surface of the mechanical structure. This difference is given by the field gradient along the z-direction times the thickness of the SiN layer. As a change in this thickness only has a negligible influence on the electrical field distribution, $G_\OM$ is effectively only proportional to the thickness. This, however, imposes a trade-off for the thickness of the mechanical resonator since a thinner layer is beneficial for the mechanical quality factor~\cite{Unterreithmeier2010,Ghadimi2018}.

\begin{figure}[!t]
    \centering
    \includegraphics[width=1.\columnwidth]{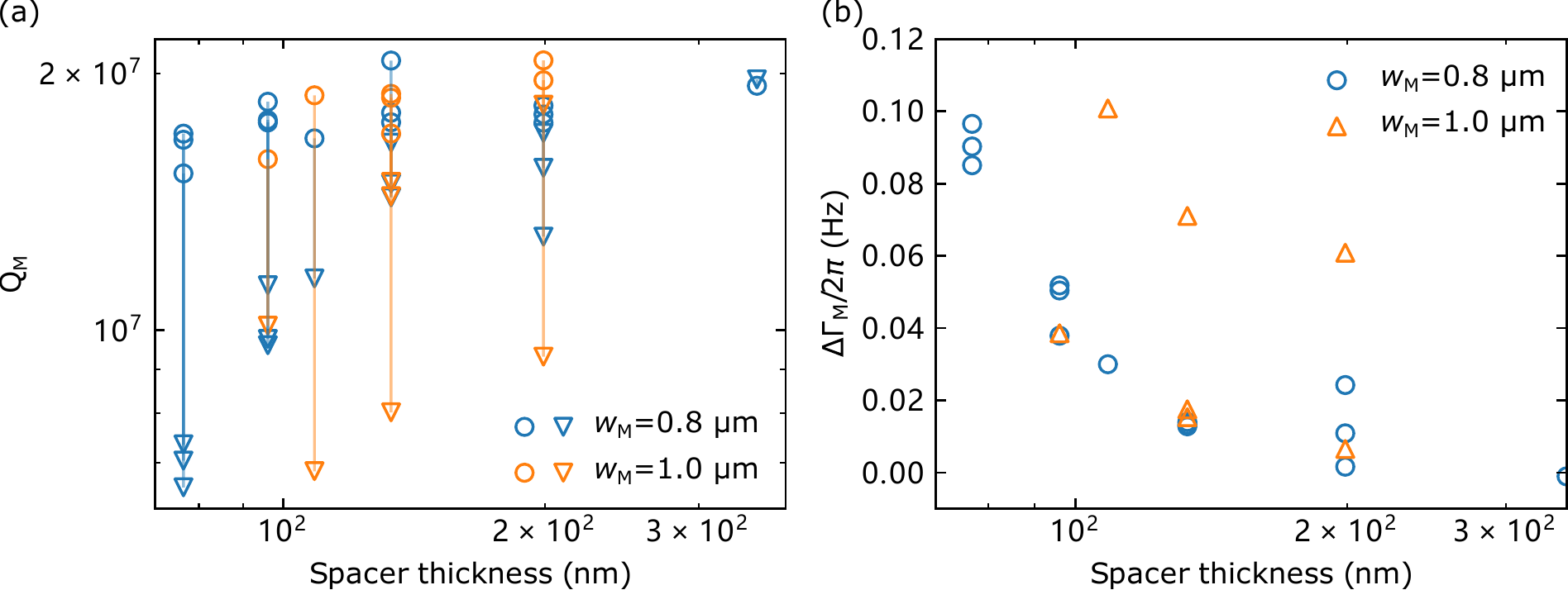}
    \caption{(a) Measured mechanical quality factor before (circle) and after (triangle) device assembly. (b) Dependence of the mechanical damping rate $\Gamma_\Mech$ on the spacer thickness.}
    \label{fig:Q_gapsize}
\end{figure}

After assembling the whole structure, we see a slight reduction in the mechanical quality factor compared to the bare resonator, which we independently measure before each integration. Interestingly, this reduction is more pronounced for smaller gaps (Figure~\ref{fig:Q_gapsize}). In order to avoid any spurious optomechanical effects in this measurement, we use large laser detunings to determine $Q_\Mech$. We therefore attribute the reduction in quality factor to the coupling of the motion of the mechanical to the optical structure. While the exact mechanism is still under investigation (e.g.\ electrostatic or van der Waals forces), we can see that by increasing the gap size the reduction of the quality factor also becomes smaller. An analysis (see Supplementary Information) shows that this reduction depends on the frequency and the quality factor of the mechanical modes of the photonic crystal. As the PhC cavity layer is stress-released, its quality factor is in general relatively low (around 500), which is why we design its frequency to be as high as possible (cf.\ Figure~\ref{fig:MechSpectMechModes}(c)).

\begin{figure}[!t]
	\centering
	\includegraphics[width=1.\columnwidth]{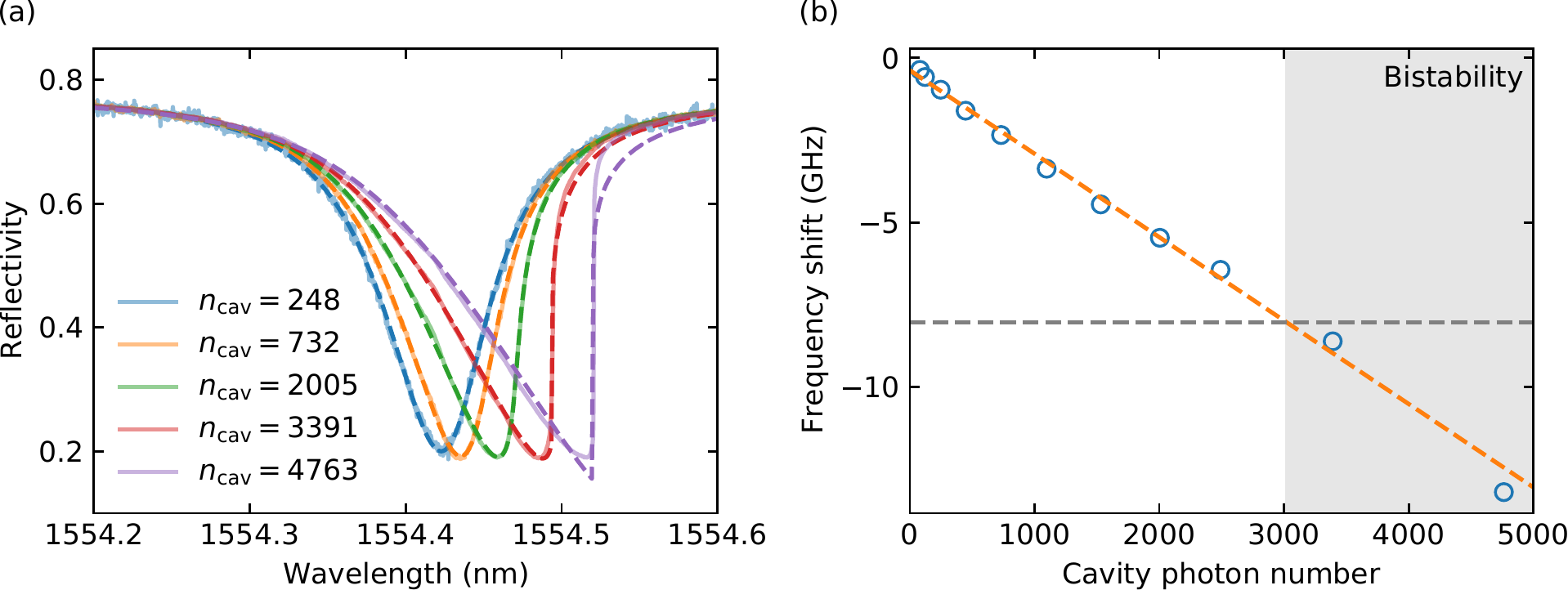}
	\caption{Optical resonance frequency shift for different input power (converted to cavity photon numbers on resonance). (a) Reflection spectrum when sweeping the laser from short wavelength to long wavelength. Dashed lines are fits to extract the frequency shift, and the results are plotted in (b). A linear fit is performed to extract the coefficient for the optical resonance tuning, which is used to obtain the bistability bound. The gray dashed line marks the maximal frequency shift above which optical bistability occurs.}
	\label{fig:ThermalTun}
\end{figure}

Being able to use a large intra-cavity photon number $n_\cav$ is important in (quantum) optomechanics, as the single photon coupling rate is enhanced as $g = \sqrt{n_\cav} g_0$. Especially, the quantum cooperativity $C_\mathrm{qu} = 4 \frac{n_\cav g_0^2}{\kappa \Gamma_\Mech n_\mathrm{th}}$, where $\Gamma_\Mech$ is the mechanical dissipation rate, $\kappa$ is the photon decay rate, and $n_\mathrm{th}$ is the phonon bath number, is a figure of merit for optomechanics in the quantum regime~\cite{Aspelmeyer2014}. It compares the photon-phonon interaction rate to the decoherence of the system, and a value comparable or even higher than unity is required for many experiments~\cite{Aspelmeyer2014,Bowen2015}. Photon absorption and the static optomechanical interaction decreases the optical resonance frequency~\cite{Eichenfield2009a}, which results in an optical bistability, for sufficiently large photon numbers~\cite{Almeida2004}. This directly limits the achievable quantum cooperativity~\cite{Aspelmeyer2014,Krause2015,Wang2019b,Jiang2020}. By introducing the additional clamping tethers on the photonic crystal we also increase the thermal anchoring to the environment and show that it is possible to achieve large $n_\cav$ before entering the bistability regime. We slowly sweep the laser across the optical resonance from short to long wavelengths at various input powers and measure the reflection. The measurements are shown in Figure~\ref{fig:ThermalTun}, for a device with $\kappa/2\pi=10.1$~GHz, gap size 130~nm and $Q_\Mech=1.49\times 10^7$. As the input power is increased, the reflection signal becomes more and more asymmetric. We fit the reflection curves and, from the asymmetry, extract the cavity resonance frequency shift. For this device, we find that the bistability occurs for $n_\cav \gtrsim 3000$, corresponding to $C_\mathrm{qu} \approx 0.2$ at room temperature, which is right around the regime required for quantum experiments~\cite{Aspelmeyer2014,Bowen2015} and several orders of magnitude higher than in previous experiments~\cite{Krause2015,Wilson2015,Guo2019}. While typically the mechanical quality increases significantly when lowering the temperature~\cite{Southworth2009}, already conservatively assuming no change in $Q_\Mech$, operating at 77~K will lead to $C_\mathrm{qu} \approx 1$.

In conclusion, we have demonstrated a simple yet highly versatile technique to integrate a photonic crystal cavity with a mechanical device, realizing large optomechanical coupling between the high-$Q_\Mech$ out-of-plane fundamental mechanical mode and the optical read-out field. This is achieved by picking spacers and photonic crystals from two separate chips, and by placing them onto the chip with the mechanical resonator. The process is robust against misalignment, while the optomechanical coupling strength is sensitive to the distance between the photonic crystal and the mechanical layer, which we can easily be set by using spacers with the desired thickness. Interestingly, the gap size influences the mechanical quality factor, which likely stems from non-optical coupling between the mechanical motion of the two layers. Our new approach paves the way to integrate almost any mechanical design with an on-chip efficient optical read-out scheme. In our demonstration, we use a system with large mechanical quality and a clean mode spectrum, which makes it highly interesting for various applications~\cite{Muhonen2019,Guo2019,Gut2020,Galinskiy2020}. Furthermore, we demonstrate that, with our newly designed photonic crystals with clamping tethers, a large intra-cavity photon number can be achieved, leading to close to unity quantum cooperativity at room temperatures. This will allow for realizing quantum optomechanical experiments and practical quantum sensor applications with no, or only very modest, cryogenic pre-cooling.

\begin{acknowledgments}
	We would like to thank Jin Chang, Maxwell Drimmer, Matthijs de Jong, Igor Marinkovi\'{c}, and Xiong Yao for valuable discussions and support. We also acknowledge assistance from the Kavli Nanolab Delft. This work is supported by the European Research Council (ERC StG Strong-Q, 676842 and ERC CoG Q-ECHOS, 101001005), and by the Netherlands Organization for Scientific Research (NWO/OCW), as part of the Frontiers of Nanoscience program, as well as through Vidi (680-47-541/994) and Vrij Programma (680-92-18-04) grants. J.G.\ gratefully acknowledges support through a Casimir PhD fellowship.
\end{acknowledgments}

\textbf{Data Availability:}\ Source data for the plots will be available on \href{}{Zenodo}.


\begin{thebibliography}{46}%
	\makeatletter
	\providecommand \@ifxundefined [1]{%
		\@ifx{#1\undefined}
	}%
	\providecommand \@ifnum [1]{%
		\ifnum #1\expandafter \@firstoftwo
		\else \expandafter \@secondoftwo
		\fi
	}%
	\providecommand \@ifx [1]{%
		\ifx #1\expandafter \@firstoftwo
		\else \expandafter \@secondoftwo
		\fi
	}%
	\providecommand \natexlab [1]{#1}%
	\providecommand \enquote  [1]{``#1''}%
	\providecommand \bibnamefont  [1]{#1}%
	\providecommand \bibfnamefont [1]{#1}%
	\providecommand \citenamefont [1]{#1}%
	\providecommand \href@noop [0]{\@secondoftwo}%
	\providecommand \href [0]{\begingroup \@sanitize@url \@href}%
	\providecommand \@href[1]{\@@startlink{#1}\@@href}%
	\providecommand \@@href[1]{\endgroup#1\@@endlink}%
	\providecommand \@sanitize@url [0]{\catcode `\\12\catcode `\$12\catcode
		`\&12\catcode `\#12\catcode `\^12\catcode `\_12\catcode `\%12\relax}%
	\providecommand \@@startlink[1]{}%
	\providecommand \@@endlink[0]{}%
	\providecommand \url  [0]{\begingroup\@sanitize@url \@url }%
	\providecommand \@url [1]{\endgroup\@href {#1}{\urlprefix }}%
	\providecommand \urlprefix  [0]{URL }%
	\providecommand \Eprint [0]{\href }%
	\providecommand \doibase [0]{https://doi.org/}%
	\providecommand \selectlanguage [0]{\@gobble}%
	\providecommand \bibinfo  [0]{\@secondoftwo}%
	\providecommand \bibfield  [0]{\@secondoftwo}%
	\providecommand \translation [1]{[#1]}%
	\providecommand \BibitemOpen [0]{}%
	\providecommand \bibitemStop [0]{}%
	\providecommand \bibitemNoStop [0]{.\EOS\space}%
	\providecommand \EOS [0]{\spacefactor3000\relax}%
	\providecommand \BibitemShut  [1]{\csname bibitem#1\endcsname}%
	\let\auto@bib@innerbib\@empty
	\bibitem [{\citenamefont {Krause}\ \emph {et~al.}(2012)\citenamefont {Krause},
		\citenamefont {Winger}, \citenamefont {Blasius}, \citenamefont {Lin},\ and\
		\citenamefont {Painter}}]{Krause2012}%
	\BibitemOpen
	\bibfield  {author} {\bibinfo {author} {\bibfnamefont {A.~G.}\ \bibnamefont
			{Krause}}, \bibinfo {author} {\bibfnamefont {M.}~\bibnamefont {Winger}},
		\bibinfo {author} {\bibfnamefont {T.~D.}\ \bibnamefont {Blasius}}, \bibinfo
		{author} {\bibfnamefont {Q.}~\bibnamefont {Lin}},\ and\ \bibinfo {author}
		{\bibfnamefont {O.}~\bibnamefont {Painter}},\ }\bibfield  {title} {\bibinfo
		{title} {A high-resolution microchip optomechanical accelerometer},\ }\href
	{https://doi.org/10.1038/NPHOTON.2012.245} {\bibfield  {journal} {\bibinfo
			{journal} {Nature Photon.}\ }\textbf {\bibinfo {volume} {6}},\ \bibinfo
		{pages} {768} (\bibinfo {year} {2012})}\BibitemShut {NoStop}%
	\bibitem [{\citenamefont {Allain}\ \emph {et~al.}(2020)\citenamefont {Allain},
		\citenamefont {Schwab}, \citenamefont {Mismer}, \citenamefont {Gely},
		\citenamefont {Mairiaux}, \citenamefont {Hermouet}, \citenamefont {Walter},
		\citenamefont {Leo}, \citenamefont {Hentz}, \citenamefont {Faucher},
		\citenamefont {Jourdan}, \citenamefont {Legrand},\ and\ \citenamefont
		{Favero}}]{Allain2020}%
	\BibitemOpen
	\bibfield  {author} {\bibinfo {author} {\bibfnamefont {P.~E.}\ \bibnamefont
			{Allain}}, \bibinfo {author} {\bibfnamefont {L.}~\bibnamefont {Schwab}},
		\bibinfo {author} {\bibfnamefont {C.}~\bibnamefont {Mismer}}, \bibinfo
		{author} {\bibfnamefont {M.}~\bibnamefont {Gely}}, \bibinfo {author}
		{\bibfnamefont {E.}~\bibnamefont {Mairiaux}}, \bibinfo {author}
		{\bibfnamefont {M.}~\bibnamefont {Hermouet}}, \bibinfo {author}
		{\bibfnamefont {B.}~\bibnamefont {Walter}}, \bibinfo {author} {\bibfnamefont
			{G.}~\bibnamefont {Leo}}, \bibinfo {author} {\bibfnamefont {S.}~\bibnamefont
			{Hentz}}, \bibinfo {author} {\bibfnamefont {M.}~\bibnamefont {Faucher}},
		\bibinfo {author} {\bibfnamefont {G.}~\bibnamefont {Jourdan}}, \bibinfo
		{author} {\bibfnamefont {B.}~\bibnamefont {Legrand}},\ and\ \bibinfo {author}
		{\bibfnamefont {I.}~\bibnamefont {Favero}},\ }\bibfield  {title} {\bibinfo
		{title} {Optomechanical resonating probe for very high frequency sensing of
			atomic forces},\ }\href {https://doi.org/10.1039/C9NR09690F} {\bibfield
		{journal} {\bibinfo  {journal} {Nanoscale}\ }\textbf {\bibinfo {volume}
			{12}},\ \bibinfo {pages} {2939} (\bibinfo {year} {2020})}\BibitemShut
	{NoStop}%
	\bibitem [{\citenamefont {Westerveld}\ \emph {et~al.}(2021)\citenamefont
		{Westerveld}, \citenamefont {Mahmud-Ul-Hasan}, \citenamefont {Shnaiderman},
		\citenamefont {Ntziachristos}, \citenamefont {Rottenberg}, \citenamefont
		{Severi},\ and\ \citenamefont {Rochus}}]{Westerveld2021}%
	\BibitemOpen
	\bibfield  {author} {\bibinfo {author} {\bibfnamefont {W.~J.}\ \bibnamefont
			{Westerveld}}, \bibinfo {author} {\bibfnamefont {M.}~\bibnamefont
			{Mahmud-Ul-Hasan}}, \bibinfo {author} {\bibfnamefont {R.}~\bibnamefont
			{Shnaiderman}}, \bibinfo {author} {\bibfnamefont {V.}~\bibnamefont
			{Ntziachristos}}, \bibinfo {author} {\bibfnamefont {X.}~\bibnamefont
			{Rottenberg}}, \bibinfo {author} {\bibfnamefont {S.}~\bibnamefont {Severi}},\
		and\ \bibinfo {author} {\bibfnamefont {V.}~\bibnamefont {Rochus}},\
	}\bibfield  {title} {\bibinfo {title} {Sensitive, small, broadband and
			scalable optomechanical ultrasound sensor in silicon photonics},\ }\href
	{https://doi.org/10.1038/s41566-021-00776-0} {\bibfield  {journal} {\bibinfo
			{journal} {Nature Photon.}\ }\textbf {\bibinfo {volume} {15}},\ \bibinfo
		{pages} {341} (\bibinfo {year} {2021})}\BibitemShut {NoStop}%
	\bibitem [{\citenamefont {Stannigel}\ \emph {et~al.}(2010)\citenamefont
		{Stannigel}, \citenamefont {Rabl}, \citenamefont {S\o{}rensen}, \citenamefont
		{Zoller},\ and\ \citenamefont {Lukin}}]{Stannigel2010}%
	\BibitemOpen
	\bibfield  {author} {\bibinfo {author} {\bibfnamefont {K.}~\bibnamefont
			{Stannigel}}, \bibinfo {author} {\bibfnamefont {P.}~\bibnamefont {Rabl}},
		\bibinfo {author} {\bibfnamefont {A.~S.}\ \bibnamefont {S\o{}rensen}},
		\bibinfo {author} {\bibfnamefont {P.}~\bibnamefont {Zoller}},\ and\ \bibinfo
		{author} {\bibfnamefont {M.~D.}\ \bibnamefont {Lukin}},\ }\bibfield  {title}
	{\bibinfo {title} {{Optomechanical Transducers for Long-Distance Quantum
				Communication}},\ }\href {https://doi.org/10.1103/PhysRevLett.105.220501}
	{\bibfield  {journal} {\bibinfo  {journal} {Phys.\ Rev.\ Lett.}\ }\textbf
		{\bibinfo {volume} {105}},\ \bibinfo {pages} {220501} (\bibinfo {year}
		{2010})}\BibitemShut {NoStop}%
	\bibitem [{\citenamefont {Wallucks}\ \emph {et~al.}(2020)\citenamefont
		{Wallucks}, \citenamefont {Marinkovi\'{c}}, \citenamefont {Hensen},
		\citenamefont {Stockill},\ and\ \citenamefont {Gr\"oblacher}}]{Wallucks2020}%
	\BibitemOpen
	\bibfield  {author} {\bibinfo {author} {\bibfnamefont {A.}~\bibnamefont
			{Wallucks}}, \bibinfo {author} {\bibfnamefont {I.}~\bibnamefont
			{Marinkovi\'{c}}}, \bibinfo {author} {\bibfnamefont {B.}~\bibnamefont
			{Hensen}}, \bibinfo {author} {\bibfnamefont {R.}~\bibnamefont {Stockill}},\
		and\ \bibinfo {author} {\bibfnamefont {S.}~\bibnamefont {Gr\"oblacher}},\
	}\bibfield  {title} {\bibinfo {title} {A quantum memory at telecom
			wavelengths},\ }\href {https://doi.org/10.1038/s41567-020-0891-z} {\bibfield
		{journal} {\bibinfo  {journal} {Nature Phys.}\ }\textbf {\bibinfo {volume}
			{16}},\ \bibinfo {pages} {772} (\bibinfo {year} {2020})}\BibitemShut
	{NoStop}%
	\bibitem [{\citenamefont {Fiaschi}\ \emph {et~al.}(2021)\citenamefont
		{Fiaschi}, \citenamefont {Hensen}, \citenamefont {Wallucks}, \citenamefont
		{Benevides}, \citenamefont {Li}, \citenamefont {Alegre},\ and\ \citenamefont
		{Gr\"oblacher}}]{Fiaschi2021}%
	\BibitemOpen
	\bibfield  {author} {\bibinfo {author} {\bibfnamefont {N.}~\bibnamefont
			{Fiaschi}}, \bibinfo {author} {\bibfnamefont {B.}~\bibnamefont {Hensen}},
		\bibinfo {author} {\bibfnamefont {A.}~\bibnamefont {Wallucks}}, \bibinfo
		{author} {\bibfnamefont {R.}~\bibnamefont {Benevides}}, \bibinfo {author}
		{\bibfnamefont {J.}~\bibnamefont {Li}}, \bibinfo {author} {\bibfnamefont
			{T.~P.~M.}\ \bibnamefont {Alegre}},\ and\ \bibinfo {author} {\bibfnamefont
			{S.}~\bibnamefont {Gr\"oblacher}},\ }\bibfield  {title} {\bibinfo {title}
		{Optomechanical quantum teleportation},\ }\href
	{https://doi.org/10.1038/s41566-021-00866-z} {\bibfield  {journal} {\bibinfo
			{journal} {Nature Photon.}\ }\textbf {\bibinfo {volume} {15}},\ \bibinfo
		{pages} {817} (\bibinfo {year} {2021})}\BibitemShut {NoStop}%
	\bibitem [{\citenamefont {Bahrami}\ \emph {et~al.}(2014)\citenamefont
		{Bahrami}, \citenamefont {Paternostro}, \citenamefont {Bassi},\ and\
		\citenamefont {Ulbricht}}]{Bahrami2014}%
	\BibitemOpen
	\bibfield  {author} {\bibinfo {author} {\bibfnamefont {M.}~\bibnamefont
			{Bahrami}}, \bibinfo {author} {\bibfnamefont {M.}~\bibnamefont
			{Paternostro}}, \bibinfo {author} {\bibfnamefont {A.}~\bibnamefont {Bassi}},\
		and\ \bibinfo {author} {\bibfnamefont {H.}~\bibnamefont {Ulbricht}},\
	}\bibfield  {title} {\bibinfo {title} {Proposal for a {Noninterferometric}
			{Test} of {Collapse} {Models} in {Optomechanical} {Systems}},\ }\href
	{https://doi.org/10.1103/PhysRevLett.112.210404} {\bibfield  {journal}
		{\bibinfo  {journal} {Phys.\ Rev.\ Lett.}\ }\textbf {\bibinfo {volume}
			{112}},\ \bibinfo {pages} {210404} (\bibinfo {year} {2014})}\BibitemShut
	{NoStop}%
	\bibitem [{\citenamefont {Carlesso}\ and\ \citenamefont
		{Donadi}(2019)}]{Carlesso2019}%
	\BibitemOpen
	\bibfield  {author} {\bibinfo {author} {\bibfnamefont {M.}~\bibnamefont
			{Carlesso}}\ and\ \bibinfo {author} {\bibfnamefont {S.}~\bibnamefont
			{Donadi}},\ }\bibfield  {title} {\bibinfo {title} {Collapse {Models}: {Main}
			{Properties} and the {State} of {Art} of the {Experimental} {Tests}},\ }in\
	\href {https://doi.org/10.1007/978-3-030-31146-9_1} {\emph {\bibinfo
			{booktitle} {Advances in {Open} {Systems} and {Fundamental} {Tests} of
				{Quantum} {Mechanics}}}},\ \bibinfo {series and number} {Springer
		{Proceedings} in {Physics}},\ \bibinfo {editor} {edited by\ \bibinfo {editor}
		{\bibfnamefont {B.}~\bibnamefont {Vacchini}}, \bibinfo {editor}
		{\bibfnamefont {H.-P.}\ \bibnamefont {Breuer}},\ and\ \bibinfo {editor}
		{\bibfnamefont {A.}~\bibnamefont {Bassi}}}\ (\bibinfo  {publisher} {Springer
		International Publishing},\ \bibinfo {year} {2019})\ pp.\ \bibinfo {pages}
	{1--13}\BibitemShut {NoStop}%
	\bibitem [{\citenamefont {Aspelmeyer}\ \emph {et~al.}(2014)\citenamefont
		{Aspelmeyer}, \citenamefont {Kippenberg},\ and\ \citenamefont
		{Marquardt}}]{Aspelmeyer2014}%
	\BibitemOpen
	\bibfield  {author} {\bibinfo {author} {\bibfnamefont {M.}~\bibnamefont
			{Aspelmeyer}}, \bibinfo {author} {\bibfnamefont {T.~J.}\ \bibnamefont
			{Kippenberg}},\ and\ \bibinfo {author} {\bibfnamefont {F.}~\bibnamefont
			{Marquardt}},\ }\bibfield  {title} {\bibinfo {title} {Cavity optomechanics},\
	}\href {https://doi.org/10.1103/RevModPhys.86.1391} {\bibfield  {journal}
		{\bibinfo  {journal} {Rev.\ Mod.\ Phys.}\ }\textbf {\bibinfo {volume} {86}},\
		\bibinfo {pages} {1391} (\bibinfo {year} {2014})}\BibitemShut {NoStop}%
	\bibitem [{\citenamefont {Norte}\ \emph {et~al.}(2016)\citenamefont {Norte},
		\citenamefont {Moura},\ and\ \citenamefont {Gr\"oblacher}}]{Norte2016}%
	\BibitemOpen
	\bibfield  {author} {\bibinfo {author} {\bibfnamefont {R.~A.}\ \bibnamefont
			{Norte}}, \bibinfo {author} {\bibfnamefont {J.~P.}\ \bibnamefont {Moura}},\
		and\ \bibinfo {author} {\bibfnamefont {S.}~\bibnamefont {Gr\"oblacher}},\
	}\bibfield  {title} {\bibinfo {title} {Mechanical resonators for quantum
			optomechanics experiments at room temperature},\ }\href
	{https://doi.org/10.1103/PhysRevLett.116.147202} {\bibfield  {journal}
		{\bibinfo  {journal} {Phys.\ Rev.\ Lett.}\ }\textbf {\bibinfo {volume}
			{116}},\ \bibinfo {pages} {147202} (\bibinfo {year} {2016})}\BibitemShut
	{NoStop}%
	\bibitem [{\citenamefont {Tsaturyan}\ \emph {et~al.}(2017)\citenamefont
		{Tsaturyan}, \citenamefont {Barg}, \citenamefont {Polzik},\ and\
		\citenamefont {Schliesser}}]{Tsaturyan2017}%
	\BibitemOpen
	\bibfield  {author} {\bibinfo {author} {\bibfnamefont {Y.}~\bibnamefont
			{Tsaturyan}}, \bibinfo {author} {\bibfnamefont {A.}~\bibnamefont {Barg}},
		\bibinfo {author} {\bibfnamefont {E.~S.}\ \bibnamefont {Polzik}},\ and\
		\bibinfo {author} {\bibfnamefont {A.}~\bibnamefont {Schliesser}},\ }\bibfield
	{title} {\bibinfo {title} {Ultracoherent nanomechanical resonators via soft
			clamping and dissipation dilution},\ }\href
	{https://doi.org/10.1038/nnano.2017.101} {\bibfield  {journal} {\bibinfo
			{journal} {Nature Nanotechn.}\ }\textbf {\bibinfo {volume} {12}},\ \bibinfo
		{pages} {776} (\bibinfo {year} {2017})}\BibitemShut {NoStop}%
	\bibitem [{\citenamefont {Ghadimi}\ \emph {et~al.}(2018)\citenamefont
		{Ghadimi}, \citenamefont {Fedorov}, \citenamefont {Engelsen}, \citenamefont
		{Bereyhi}, \citenamefont {Schilling}, \citenamefont {Wilson},\ and\
		\citenamefont {Kippenberg}}]{Ghadimi2018}%
	\BibitemOpen
	\bibfield  {author} {\bibinfo {author} {\bibfnamefont {A.~H.}\ \bibnamefont
			{Ghadimi}}, \bibinfo {author} {\bibfnamefont {S.~A.}\ \bibnamefont
			{Fedorov}}, \bibinfo {author} {\bibfnamefont {N.~J.}\ \bibnamefont
			{Engelsen}}, \bibinfo {author} {\bibfnamefont {M.~J.}\ \bibnamefont
			{Bereyhi}}, \bibinfo {author} {\bibfnamefont {R.}~\bibnamefont {Schilling}},
		\bibinfo {author} {\bibfnamefont {D.~J.}\ \bibnamefont {Wilson}},\ and\
		\bibinfo {author} {\bibfnamefont {T.~J.}\ \bibnamefont {Kippenberg}},\
	}\bibfield  {title} {\bibinfo {title} {Elastic strain engineering for
			ultralow mechanical dissipation},\ }\href
	{https://doi.org/10.1126/science.aar6939} {\bibfield  {journal} {\bibinfo
			{journal} {Science}\ }\textbf {\bibinfo {volume} {360}},\ \bibinfo {pages}
		{764} (\bibinfo {year} {2018})}\BibitemShut {NoStop}%
	\bibitem [{\citenamefont {H{\o}j}\ \emph {et~al.}(2021)\citenamefont {H{\o}j},
		\citenamefont {Wang}, \citenamefont {Gao}, \citenamefont {Hoff},
		\citenamefont {Sigmund},\ and\ \citenamefont {Andersen}}]{Hoej2021}%
	\BibitemOpen
	\bibfield  {author} {\bibinfo {author} {\bibfnamefont {D.}~\bibnamefont
			{H{\o}j}}, \bibinfo {author} {\bibfnamefont {F.}~\bibnamefont {Wang}},
		\bibinfo {author} {\bibfnamefont {W.}~\bibnamefont {Gao}}, \bibinfo {author}
		{\bibfnamefont {U.~B.}\ \bibnamefont {Hoff}}, \bibinfo {author}
		{\bibfnamefont {O.}~\bibnamefont {Sigmund}},\ and\ \bibinfo {author}
		{\bibfnamefont {U.~L.}\ \bibnamefont {Andersen}},\ }\bibfield  {title}
	{\bibinfo {title} {Ultra-coherent nanomechanical resonators based on inverse
			design},\ }\href {https://doi.org/10.1038/s41467-021-26102-4} {\bibfield
		{journal} {\bibinfo  {journal} {Nature Commun.}\ }\textbf {\bibinfo {volume}
			{12}},\ \bibinfo {pages} {5766} (\bibinfo {year} {2021})}\BibitemShut
	{NoStop}%
	\bibitem [{\citenamefont {Beccari}\ \emph {et~al.}(2021)\citenamefont
		{Beccari}, \citenamefont {Bereyhi}, \citenamefont {Groth}, \citenamefont
		{Fedorov}, \citenamefont {Arabmoheghi}, \citenamefont {Engelsen},\ and\
		\citenamefont {Kippenberg}}]{Beccari2021}%
	\BibitemOpen
	\bibfield  {author} {\bibinfo {author} {\bibfnamefont {A.}~\bibnamefont
			{Beccari}}, \bibinfo {author} {\bibfnamefont {M.~J.}\ \bibnamefont
			{Bereyhi}}, \bibinfo {author} {\bibfnamefont {R.}~\bibnamefont {Groth}},
		\bibinfo {author} {\bibfnamefont {S.~A.}\ \bibnamefont {Fedorov}}, \bibinfo
		{author} {\bibfnamefont {A.}~\bibnamefont {Arabmoheghi}}, \bibinfo {author}
		{\bibfnamefont {N.~J.}\ \bibnamefont {Engelsen}},\ and\ \bibinfo {author}
		{\bibfnamefont {T.~J.}\ \bibnamefont {Kippenberg}},\ }\bibfield  {title}
	{\bibinfo {title} {Hierarchical tensile structures with ultralow mechanical
			dissipation},\ }\href {http://arxiv.org/abs/2103.09785} {\bibfield  {journal}
		{\bibinfo  {journal} {arXiv:2103.09785}\ } (\bibinfo {year}
		{2021})}\BibitemShut {NoStop}%
	\bibitem [{\citenamefont {Norte}\ \emph {et~al.}(2018)\citenamefont {Norte},
		\citenamefont {Forsch}, \citenamefont {Wallucks}, \citenamefont
		{Marinkovi\'{c}},\ and\ \citenamefont {Gr\"oblacher}}]{Norte2018}%
	\BibitemOpen
	\bibfield  {author} {\bibinfo {author} {\bibfnamefont {R.~A.}\ \bibnamefont
			{Norte}}, \bibinfo {author} {\bibfnamefont {M.}~\bibnamefont {Forsch}},
		\bibinfo {author} {\bibfnamefont {A.}~\bibnamefont {Wallucks}}, \bibinfo
		{author} {\bibfnamefont {I.}~\bibnamefont {Marinkovi\'{c}}},\ and\ \bibinfo
		{author} {\bibfnamefont {S.}~\bibnamefont {Gr\"oblacher}},\ }\bibfield
	{title} {\bibinfo {title} {Platform for measurements of the casimir force
			between two superconductors},\ }\href
	{https://doi.org/10.1103/PhysRevLett.121.030405} {\bibfield  {journal}
		{\bibinfo  {journal} {Phys.\ Rev.\ Lett.}\ }\textbf {\bibinfo {volume}
			{121}},\ \bibinfo {pages} {030405} (\bibinfo {year} {2018})}\BibitemShut
	{NoStop}%
	\bibitem [{\citenamefont {Whittle}\ \emph {et~al.}(2021)\citenamefont
		{Whittle}, \citenamefont {Hall}, \citenamefont {Dwyer}, \citenamefont
		{Mavalvala}, \citenamefont {Sudhir}, \citenamefont {Abbott}, \citenamefont
		{Ananyeva}, \citenamefont {Austin}, \citenamefont {Barsotti}, \citenamefont
		{Betzwieser}, \citenamefont {Blair}, \citenamefont {Brooks}, \citenamefont
		{Brown}, \citenamefont {Buikema}, \citenamefont {Cahillane}, \citenamefont
		{Driggers}, \citenamefont {Effler}, \citenamefont {Fernandez-Galiana},
		\citenamefont {Fritschel}, \citenamefont {Frolov}, \citenamefont {Hardwick},
		\citenamefont {Kasprzack}, \citenamefont {Kawabe}, \citenamefont
		{Kijbunchoo}, \citenamefont {Kissel}, \citenamefont {Mansell}, \citenamefont
		{Matichard}, \citenamefont {McCuller}, \citenamefont {McRae}, \citenamefont
		{Mullavey}, \citenamefont {Pele}, \citenamefont {Schofield}, \citenamefont
		{Sigg}, \citenamefont {Tse}, \citenamefont {Vajente}, \citenamefont
		{Vander-Hyde}, \citenamefont {Yu}, \citenamefont {Yu}, \citenamefont {Adams},
		\citenamefont {Adhikari}, \citenamefont {Appert}, \citenamefont {Arai},
		\citenamefont {Areeda}, \citenamefont {Asali}, \citenamefont {Aston},
		\citenamefont {Baer}, \citenamefont {Ball}, \citenamefont {Ballmer},
		\citenamefont {Banagiri}, \citenamefont {Barker}, \citenamefont {Bartlett},
		\citenamefont {Berger}, \citenamefont {Bhattacharjee}, \citenamefont
		{Billingsley}, \citenamefont {Biscans}, \citenamefont {Blair}, \citenamefont
		{Bode}, \citenamefont {Booker}, \citenamefont {Bork}, \citenamefont
		{Bramley}, \citenamefont {Cannon}, \citenamefont {Chen}, \citenamefont
		{Ciobanu}, \citenamefont {Clara}, \citenamefont {Compton}, \citenamefont
		{Cooper}, \citenamefont {Corley}, \citenamefont {Countryman}, \citenamefont
		{Covas}, \citenamefont {Coyne}, \citenamefont {Datrier}, \citenamefont
		{Davis}, \citenamefont {Di~Fronzo}, \citenamefont {Dooley}, \citenamefont
		{Dupej}, \citenamefont {Etzel}, \citenamefont {Evans}, \citenamefont {Evans},
		\citenamefont {Feicht}, \citenamefont {Fulda}, \citenamefont {Fyffe},
		\citenamefont {Giaime}, \citenamefont {Giardina}, \citenamefont {Godwin},
		\citenamefont {Goetz}, \citenamefont {Gras}, \citenamefont {Gray},
		\citenamefont {Gray}, \citenamefont {Green}, \citenamefont {Gustafson},
		\citenamefont {Gustafson}, \citenamefont {Hanks}, \citenamefont {Hanson},
		\citenamefont {Hasskew}, \citenamefont {Heintze}, \citenamefont
		{Helmling-Cornell}, \citenamefont {Holland}, \citenamefont {Jones},
		\citenamefont {Kandhasamy}, \citenamefont {Karki}, \citenamefont {King},
		\citenamefont {Kumar}, \citenamefont {Landry}, \citenamefont {Lane},
		\citenamefont {Lantz}, \citenamefont {Laxen}, \citenamefont {Lecoeuche},
		\citenamefont {Leviton}, \citenamefont {Liu}, \citenamefont {Lormand},
		\citenamefont {Lundgren}, \citenamefont {Macas}, \citenamefont {MacInnis},
		\citenamefont {Macleod}, \citenamefont {M{\'a}rka}, \citenamefont
		{M{\'a}rka}, \citenamefont {Martynov}, \citenamefont {Mason}, \citenamefont
		{Massinger}, \citenamefont {McCarthy}, \citenamefont {McClelland},
		\citenamefont {McCormick}, \citenamefont {McIver}, \citenamefont {Mendell},
		\citenamefont {Merfeld}, \citenamefont {Merilh}, \citenamefont {Meylahn},
		\citenamefont {Mistry}, \citenamefont {Mittleman}, \citenamefont {Moreno},
		\citenamefont {Mow-Lowry}, \citenamefont {Mozzon}, \citenamefont {Nelson},
		\citenamefont {Nguyen}, \citenamefont {Nuttall}, \citenamefont {Oberling},
		\citenamefont {Oram}, \citenamefont {Osthelder}, \citenamefont {Ottaway},
		\citenamefont {Overmier}, \citenamefont {Palamos}, \citenamefont {Parker},
		\citenamefont {Payne}, \citenamefont {Penhorwood}, \citenamefont {Perez},
		\citenamefont {Pirello}, \citenamefont {Radkins}, \citenamefont {Ramirez},
		\citenamefont {Richardson}, \citenamefont {Riles}, \citenamefont {Robertson},
		\citenamefont {Rollins}, \citenamefont {Romel}, \citenamefont {Romie},
		\citenamefont {Ross}, \citenamefont {Ryan}, \citenamefont {Sadecki},
		\citenamefont {Sanchez}, \citenamefont {Sanchez}, \citenamefont {Saravanan},
		\citenamefont {Savage}, \citenamefont {Schaetz}, \citenamefont {Schnabel},
		\citenamefont {Schwartz}, \citenamefont {Sellers}, \citenamefont {Shaffer},
		\citenamefont {Slagmolen}, \citenamefont {Smith}, \citenamefont {Soni},
		\citenamefont {Sorazu}, \citenamefont {Spencer}, \citenamefont {Strain},
		\citenamefont {Sun}, \citenamefont {Szczepa{\'n}czyk}, \citenamefont
		{Thomas}, \citenamefont {Thomas}, \citenamefont {Thorne}, \citenamefont
		{Toland}, \citenamefont {Torrie}, \citenamefont {Traylor}, \citenamefont
		{Urban}, \citenamefont {Valdes}, \citenamefont {Veitch}, \citenamefont
		{Venkateswara}, \citenamefont {Venugopalan}, \citenamefont {Viets},
		\citenamefont {Vo}, \citenamefont {Vorvick}, \citenamefont {Wade},
		\citenamefont {Ward}, \citenamefont {Warner}, \citenamefont {Weaver},
		\citenamefont {Weiss}, \citenamefont {Willke}, \citenamefont {Wipf},
		\citenamefont {Xiao}, \citenamefont {Yamamoto}, \citenamefont {Zhang},
		\citenamefont {Zucker},\ and\ \citenamefont {Zweizig}}]{Whittle2021}%
	\BibitemOpen
	\bibfield  {author} {\bibinfo {author} {\bibfnamefont {C.}~\bibnamefont
			{Whittle}}, \bibinfo {author} {\bibfnamefont {et}\ \bibnamefont {al.}},\ }\bibfield  {title}
	{\bibinfo {title} {Approaching the motional ground state of a 10-kg object},\
	}\href {https://doi.org/10.1126/science.abh2634} {\bibfield  {journal}
		{\bibinfo  {journal} {Science}\ }\textbf {\bibinfo {volume} {372}},\ \bibinfo
		{pages} {1333} (\bibinfo {year} {2021})}\BibitemShut {NoStop}%
	\bibitem [{\citenamefont {Gut}\ \emph {et~al.}(2020)\citenamefont {Gut},
		\citenamefont {Winkler}, \citenamefont {Hoelscher-Obermaier}, \citenamefont
		{Hofer}, \citenamefont {Nia}, \citenamefont {Walk}, \citenamefont {Steffens},
		\citenamefont {Eisert}, \citenamefont {Wieczorek}, \citenamefont {Slater},
		\citenamefont {Aspelmeyer},\ and\ \citenamefont {Hammerer}}]{Gut2020}%
	\BibitemOpen
	\bibfield  {author} {\bibinfo {author} {\bibfnamefont {C.}~\bibnamefont
			{Gut}}, \bibinfo {author} {\bibfnamefont {K.}~\bibnamefont {Winkler}},
		\bibinfo {author} {\bibfnamefont {J.}~\bibnamefont {Hoelscher-Obermaier}},
		\bibinfo {author} {\bibfnamefont {S.~G.}\ \bibnamefont {Hofer}}, \bibinfo
		{author} {\bibfnamefont {R.~M.}\ \bibnamefont {Nia}}, \bibinfo {author}
		{\bibfnamefont {N.}~\bibnamefont {Walk}}, \bibinfo {author} {\bibfnamefont
			{A.}~\bibnamefont {Steffens}}, \bibinfo {author} {\bibfnamefont
			{J.}~\bibnamefont {Eisert}}, \bibinfo {author} {\bibfnamefont
			{W.}~\bibnamefont {Wieczorek}}, \bibinfo {author} {\bibfnamefont {J.~A.}\
			\bibnamefont {Slater}}, \bibinfo {author} {\bibfnamefont {M.}~\bibnamefont
			{Aspelmeyer}},\ and\ \bibinfo {author} {\bibfnamefont {K.}~\bibnamefont
			{Hammerer}},\ }\bibfield  {title} {\bibinfo {title} {Stationary
			optomechanical entanglement between a mechanical oscillator and its
			measurement apparatus},\ }\href
	{https://doi.org/10.1103/PhysRevResearch.2.033244} {\bibfield  {journal}
		{\bibinfo  {journal} {Phys.\ Rev.\ Research}\ }\textbf {\bibinfo {volume}
			{2}},\ \bibinfo {pages} {033244} (\bibinfo {year} {2020})}\BibitemShut
	{NoStop}%
	\bibitem [{\citenamefont {Wilson}\ \emph {et~al.}(2015)\citenamefont {Wilson},
		\citenamefont {Sudhir}, \citenamefont {Piro}, \citenamefont {Schilling},
		\citenamefont {Ghadimi},\ and\ \citenamefont {Kippenberg}}]{Wilson2015}%
	\BibitemOpen
	\bibfield  {author} {\bibinfo {author} {\bibfnamefont {D.~J.}\ \bibnamefont
			{Wilson}}, \bibinfo {author} {\bibfnamefont {V.}~\bibnamefont {Sudhir}},
		\bibinfo {author} {\bibfnamefont {N.}~\bibnamefont {Piro}}, \bibinfo {author}
		{\bibfnamefont {R.}~\bibnamefont {Schilling}}, \bibinfo {author}
		{\bibfnamefont {A.}~\bibnamefont {Ghadimi}},\ and\ \bibinfo {author}
		{\bibfnamefont {T.~J.}\ \bibnamefont {Kippenberg}},\ }\bibfield  {title}
	{\bibinfo {title} {Measurement-based control of a mechanical oscillator at
			its thermal decoherence rate},\ }\href {https://doi.org/10.1038/nature14672}
	{\bibfield  {journal} {\bibinfo  {journal} {Nature}\ }\textbf {\bibinfo
			{volume} {524}},\ \bibinfo {pages} {325} (\bibinfo {year}
		{2015})}\BibitemShut {NoStop}%
	\bibitem [{\citenamefont {Unterreithmeier}\ \emph {et~al.}(2010)\citenamefont
		{Unterreithmeier}, \citenamefont {Faust},\ and\ \citenamefont
		{Kotthaus}}]{Unterreithmeier2010}%
	\BibitemOpen
	\bibfield  {author} {\bibinfo {author} {\bibfnamefont {Q.~P.}\ \bibnamefont
			{Unterreithmeier}}, \bibinfo {author} {\bibfnamefont {T.}~\bibnamefont
			{Faust}},\ and\ \bibinfo {author} {\bibfnamefont {J.~P.}\ \bibnamefont
			{Kotthaus}},\ }\bibfield  {title} {\bibinfo {title} {Damping of
			nanomechanical resonators},\ }\href
	{https://doi.org/10.1103/PhysRevLett.105.027205} {\bibfield  {journal}
		{\bibinfo  {journal} {Phys.\ Rev.\ Lett.}\ }\textbf {\bibinfo {volume}
			{105}},\ \bibinfo {pages} {027205} (\bibinfo {year} {2010})}\BibitemShut
	{NoStop}%
	\bibitem [{\citenamefont {Fedorov}\ \emph {et~al.}(2020)\citenamefont
		{Fedorov}, \citenamefont {Beccari}, \citenamefont {Engelsen},\ and\
		\citenamefont {Kippenberg}}]{Fedorov2020}%
	\BibitemOpen
	\bibfield  {author} {\bibinfo {author} {\bibfnamefont {S.}~\bibnamefont
			{Fedorov}}, \bibinfo {author} {\bibfnamefont {A.}~\bibnamefont {Beccari}},
		\bibinfo {author} {\bibfnamefont {N.}~\bibnamefont {Engelsen}},\ and\
		\bibinfo {author} {\bibfnamefont {T.}~\bibnamefont {Kippenberg}},\ }\bibfield
	{title} {\bibinfo {title} {Fractal-like mechanical resonators with a
			soft-clamped fundamental mode},\ }\href
	{https://doi.org/10.1103/physrevlett.124.025502} {\bibfield  {journal}
		{\bibinfo  {journal} {Phys.\ Rev.\ Lett.}\ }\textbf {\bibinfo {volume}
			{124}},\ \bibinfo {pages} {025502} (\bibinfo {year} {2020})}\BibitemShut
	{NoStop}%
	\bibitem [{\citenamefont {Albrecht}\ \emph {et~al.}(1991)\citenamefont
		{Albrecht}, \citenamefont {Gr\"utter}, \citenamefont {Horne},\ and\
		\citenamefont {Rugar}}]{Albrecht1991}%
	\BibitemOpen
	\bibfield  {author} {\bibinfo {author} {\bibfnamefont {T.~R.}\ \bibnamefont
			{Albrecht}}, \bibinfo {author} {\bibfnamefont {P.}~\bibnamefont {Gr\"utter}},
		\bibinfo {author} {\bibfnamefont {D.}~\bibnamefont {Horne}},\ and\ \bibinfo
		{author} {\bibfnamefont {D.}~\bibnamefont {Rugar}},\ }\bibfield  {title}
	{\bibinfo {title} {Frequency modulation detection using high-{Q} cantilevers
			for enhanced force microscope sensitivity},\ }\href
	{https://doi.org/10.1063/1.347347} {\bibfield  {journal} {\bibinfo  {journal}
			{J.\ Appl.\ Phys.}\ }\textbf {\bibinfo {volume} {69}},\ \bibinfo {pages}
		{668} (\bibinfo {year} {1991})}\BibitemShut {NoStop}%
	\bibitem [{\citenamefont {Carlesso}\ \emph {et~al.}(2018)\citenamefont
		{Carlesso}, \citenamefont {Vinante},\ and\ \citenamefont
		{Bassi}}]{Carlesso2018}%
	\BibitemOpen
	\bibfield  {author} {\bibinfo {author} {\bibfnamefont {M.}~\bibnamefont
			{Carlesso}}, \bibinfo {author} {\bibfnamefont {A.}~\bibnamefont {Vinante}},\
		and\ \bibinfo {author} {\bibfnamefont {A.}~\bibnamefont {Bassi}},\ }\bibfield
	{title} {\bibinfo {title} {Multilayer test masses to enhance the collapse
			noise},\ }\href {https://doi.org/10.1103/PhysRevA.98.022122} {\bibfield
		{journal} {\bibinfo  {journal} {Phys.\ Rev.\ A}\ }\textbf {\bibinfo {volume}
			{98}},\ \bibinfo {pages} {022122} (\bibinfo {year} {2018})}\BibitemShut
	{NoStop}%
	\bibitem [{\citenamefont {Pate}\ \emph {et~al.}(2020)\citenamefont {Pate},
		\citenamefont {Goryachev}, \citenamefont {Chiao}, \citenamefont {Sharping},\
		and\ \citenamefont {Tobar}}]{Pate2020}%
	\BibitemOpen
	\bibfield  {author} {\bibinfo {author} {\bibfnamefont {J.~M.}\ \bibnamefont
			{Pate}}, \bibinfo {author} {\bibfnamefont {M.}~\bibnamefont {Goryachev}},
		\bibinfo {author} {\bibfnamefont {R.~Y.}\ \bibnamefont {Chiao}}, \bibinfo
		{author} {\bibfnamefont {J.~E.}\ \bibnamefont {Sharping}},\ and\ \bibinfo
		{author} {\bibfnamefont {M.~E.}\ \bibnamefont {Tobar}},\ }\bibfield  {title}
	{\bibinfo {title} {Casimir spring and dilution in macroscopic cavity
			optomechanics},\ }\href {https://doi.org/10.1038/s41567-020-0975-9}
	{\bibfield  {journal} {\bibinfo  {journal} {Nature Phys.}\ }\textbf {\bibinfo
			{volume} {16}},\ \bibinfo {pages} {1117} (\bibinfo {year}
		{2020})}\BibitemShut {NoStop}%
	\bibitem [{\citenamefont {H\"alg}\ \emph {et~al.}(2021)\citenamefont {H\"alg},
		\citenamefont {Gisler}, \citenamefont {Tsaturyan}, \citenamefont {Catalini},
		\citenamefont {Grob}, \citenamefont {Krass}, \citenamefont {H\'{e}ritier},
		\citenamefont {Mattiat}, \citenamefont {Thamm}, \citenamefont {Schirhagl},
		\citenamefont {Langman}, \citenamefont {Schliesser}, \citenamefont {Degen},\
		and\ \citenamefont {Eichler}}]{Haelg2021}%
	\BibitemOpen
	\bibfield  {author} {\bibinfo {author} {\bibfnamefont {D.}~\bibnamefont
			{H\"alg}}, \bibinfo {author} {\bibfnamefont {T.}~\bibnamefont {Gisler}},
		\bibinfo {author} {\bibfnamefont {Y.}~\bibnamefont {Tsaturyan}}, \bibinfo
		{author} {\bibfnamefont {L.}~\bibnamefont {Catalini}}, \bibinfo {author}
		{\bibfnamefont {U.}~\bibnamefont {Grob}}, \bibinfo {author} {\bibfnamefont
			{M.-D.}\ \bibnamefont {Krass}}, \bibinfo {author} {\bibfnamefont
			{M.}~\bibnamefont {H\'{e}ritier}}, \bibinfo {author} {\bibfnamefont
			{H.}~\bibnamefont {Mattiat}}, \bibinfo {author} {\bibfnamefont {A.-K.}\
			\bibnamefont {Thamm}}, \bibinfo {author} {\bibfnamefont {R.}~\bibnamefont
			{Schirhagl}}, \bibinfo {author} {\bibfnamefont {E.~C.}\ \bibnamefont
			{Langman}}, \bibinfo {author} {\bibfnamefont {A.}~\bibnamefont {Schliesser}},
		\bibinfo {author} {\bibfnamefont {C.~L.}\ \bibnamefont {Degen}},\ and\
		\bibinfo {author} {\bibfnamefont {A.}~\bibnamefont {Eichler}},\ }\bibfield
	{title} {\bibinfo {title} {Membrane-{Based} {Scanning} {Force}
			{Microscopy}},\ }\href {https://doi.org/10.1103/PhysRevApplied.15.L021001}
	{\bibfield  {journal} {\bibinfo  {journal} {Phys.\ Rev.\ Applied}\ }\textbf
		{\bibinfo {volume} {15}},\ \bibinfo {pages} {L021001} (\bibinfo {year}
		{2021})}\BibitemShut {NoStop}%
	\bibitem [{\citenamefont {Muhonen}\ \emph {et~al.}(2019)\citenamefont
		{Muhonen}, \citenamefont {Gala}, \citenamefont {Leijssen},\ and\
		\citenamefont {Verhagen}}]{Muhonen2019}%
	\BibitemOpen
	\bibfield  {author} {\bibinfo {author} {\bibfnamefont {J.~T.}\ \bibnamefont
			{Muhonen}}, \bibinfo {author} {\bibfnamefont {G.~R.~L.}\ \bibnamefont
			{Gala}}, \bibinfo {author} {\bibfnamefont {R.}~\bibnamefont {Leijssen}},\
		and\ \bibinfo {author} {\bibfnamefont {E.}~\bibnamefont {Verhagen}},\
	}\bibfield  {title} {\bibinfo {title} {State {Preparation} and {Tomography}
			of a {Nanomechanical} {Resonator} with {Fast} {Light} {Pulses}},\ }\href
	{https://doi.org/10.1103/PhysRevLett.123.113601} {\bibfield  {journal}
		{\bibinfo  {journal} {Phys.\ Rev.\ Lett.}\ }\textbf {\bibinfo {volume}
			{123}},\ \bibinfo {pages} {113601} (\bibinfo {year} {2019})}\BibitemShut
	{NoStop}%
	\bibitem [{\citenamefont {Guo}\ \emph {et~al.}(2019)\citenamefont {Guo},
		\citenamefont {Norte},\ and\ \citenamefont {Gr\"oblacher}}]{Guo2019}%
	\BibitemOpen
	\bibfield  {author} {\bibinfo {author} {\bibfnamefont {J.}~\bibnamefont
			{Guo}}, \bibinfo {author} {\bibfnamefont {R.}~\bibnamefont {Norte}},\ and\
		\bibinfo {author} {\bibfnamefont {S.}~\bibnamefont {Gr\"oblacher}},\
	}\bibfield  {title} {\bibinfo {title} {Feedback {Cooling} of a {Room}
			{Temperature} {Mechanical} {Oscillator} close to its {Motional} {Ground}
			{State}},\ }\href {https://doi.org/10.1103/PhysRevLett.123.223602} {\bibfield
		{journal} {\bibinfo  {journal} {Phys.\ Rev.\ Lett.}\ }\textbf {\bibinfo
			{volume} {123}},\ \bibinfo {pages} {223602} (\bibinfo {year}
		{2019})}\BibitemShut {NoStop}%
	\bibitem [{\citenamefont {Galinskiy}\ \emph {et~al.}(2020)\citenamefont
		{Galinskiy}, \citenamefont {Tsaturyan}, \citenamefont {Parniak},\ and\
		\citenamefont {Polzik}}]{Galinskiy2020}%
	\BibitemOpen
	\bibfield  {author} {\bibinfo {author} {\bibfnamefont {I.}~\bibnamefont
			{Galinskiy}}, \bibinfo {author} {\bibfnamefont {Y.}~\bibnamefont
			{Tsaturyan}}, \bibinfo {author} {\bibfnamefont {M.}~\bibnamefont {Parniak}},\
		and\ \bibinfo {author} {\bibfnamefont {E.~S.}\ \bibnamefont {Polzik}},\
	}\bibfield  {title} {\bibinfo {title} {Phonon counting thermometry of an
			ultracoherent membrane resonator near its motional ground state},\ }\href
	{https://doi.org/10.1364/OPTICA.390939} {\bibfield  {journal} {\bibinfo
			{journal} {Optica}\ }\textbf {\bibinfo {volume} {7}},\ \bibinfo {pages} {718}
		(\bibinfo {year} {2020})}\BibitemShut {NoStop}%
	\bibitem [{\citenamefont {\v{Z}arko Zobenica}\ \emph
		{et~al.}(2017)\citenamefont {\v{Z}arko Zobenica}, \citenamefont {van~der
			Heijden}, \citenamefont {Petruzzella}, \citenamefont {Pagliano},
		\citenamefont {Leijssen}, \citenamefont {Xia}, \citenamefont {Midolo},
		\citenamefont {Cotrufo}, \citenamefont {Cho}, \citenamefont {van Otten},
		\citenamefont {Verhagen},\ and\ \citenamefont {Fiore}}]{Zobenica2017}%
	\BibitemOpen
	\bibfield  {author} {\bibinfo {author} {\bibnamefont {\v{Z}arko Zobenica}},
		\bibinfo {author} {\bibfnamefont {R.~W.}\ \bibnamefont {van~der Heijden}},
		\bibinfo {author} {\bibfnamefont {M.}~\bibnamefont {Petruzzella}}, \bibinfo
		{author} {\bibfnamefont {F.}~\bibnamefont {Pagliano}}, \bibinfo {author}
		{\bibfnamefont {R.}~\bibnamefont {Leijssen}}, \bibinfo {author}
		{\bibfnamefont {T.}~\bibnamefont {Xia}}, \bibinfo {author} {\bibfnamefont
			{L.}~\bibnamefont {Midolo}}, \bibinfo {author} {\bibfnamefont
			{M.}~\bibnamefont {Cotrufo}}, \bibinfo {author} {\bibfnamefont
			{Y.}~\bibnamefont {Cho}}, \bibinfo {author} {\bibfnamefont {F.~W.~M.}\
			\bibnamefont {van Otten}}, \bibinfo {author} {\bibfnamefont {E.}~\bibnamefont
			{Verhagen}},\ and\ \bibinfo {author} {\bibfnamefont {A.}~\bibnamefont
			{Fiore}},\ }\bibfield  {title} {\bibinfo {title} {Integrated
			nano-opto-electro-mechanical sensor for spectrometry and nanometrology},\
	}\href {https://doi.org/10.1038/s41467-017-02392-5} {\bibfield  {journal}
		{\bibinfo  {journal} {Nature Commun.}\ }\textbf {\bibinfo {volume} {8}},\
		\bibinfo {pages} {2216} (\bibinfo {year} {2017})}\BibitemShut {NoStop}%
	\bibitem [{\citenamefont {Liu}\ \emph {et~al.}(2020)\citenamefont {Liu},
		\citenamefont {Pagliano}, \citenamefont {van Veldhoven}, \citenamefont
		{Pogoretskiy}, \citenamefont {Jiao},\ and\ \citenamefont {Fiore}}]{Liu2020}%
	\BibitemOpen
	\bibfield  {author} {\bibinfo {author} {\bibfnamefont {T.}~\bibnamefont
			{Liu}}, \bibinfo {author} {\bibfnamefont {F.}~\bibnamefont {Pagliano}},
		\bibinfo {author} {\bibfnamefont {R.}~\bibnamefont {van Veldhoven}}, \bibinfo
		{author} {\bibfnamefont {V.}~\bibnamefont {Pogoretskiy}}, \bibinfo {author}
		{\bibfnamefont {Y.}~\bibnamefont {Jiao}},\ and\ \bibinfo {author}
		{\bibfnamefont {A.}~\bibnamefont {Fiore}},\ }\bibfield  {title} {\bibinfo
		{title} {Integrated nano-optomechanical displacement sensor with ultrawide
			optical bandwidth},\ }\href {https://doi.org/10.1038/s41467-020-16269-7}
	{\bibfield  {journal} {\bibinfo  {journal} {Nature Commun.}\ }\textbf
		{\bibinfo {volume} {11}},\ \bibinfo {pages} {2407} (\bibinfo {year}
		{2020})}\BibitemShut {NoStop}%
	\bibitem [{\citenamefont {Elshaari}\ \emph {et~al.}(2020)\citenamefont
		{Elshaari}, \citenamefont {Pernice}, \citenamefont {Srinivasan},
		\citenamefont {Benson},\ and\ \citenamefont {Zwiller}}]{Elshaari2020}%
	\BibitemOpen
	\bibfield  {author} {\bibinfo {author} {\bibfnamefont {A.~W.}\ \bibnamefont
			{Elshaari}}, \bibinfo {author} {\bibfnamefont {W.}~\bibnamefont {Pernice}},
		\bibinfo {author} {\bibfnamefont {K.}~\bibnamefont {Srinivasan}}, \bibinfo
		{author} {\bibfnamefont {O.}~\bibnamefont {Benson}},\ and\ \bibinfo {author}
		{\bibfnamefont {V.}~\bibnamefont {Zwiller}},\ }\bibfield  {title} {\bibinfo
		{title} {Hybrid integrated quantum photonic circuits},\ }\href
	{https://doi.org/10.1038/s41566-020-0609-x} {\bibfield  {journal} {\bibinfo
			{journal} {Nature Photon.}\ }\textbf {\bibinfo {volume} {14}},\ \bibinfo
		{pages} {285} (\bibinfo {year} {2020})}\BibitemShut {NoStop}%
	\bibitem [{\citenamefont {Marinkovi\'c}\ \emph {et~al.}(2021)\citenamefont
		{Marinkovi\'c}, \citenamefont {Drimmer}, \citenamefont {Hensen},\ and\
		\citenamefont {Gr\"oblacher}}]{Marinkovic2021}%
	\BibitemOpen
	\bibfield  {author} {\bibinfo {author} {\bibfnamefont {I.}~\bibnamefont
			{Marinkovi\'c}}, \bibinfo {author} {\bibfnamefont {M.}~\bibnamefont
			{Drimmer}}, \bibinfo {author} {\bibfnamefont {B.}~\bibnamefont {Hensen}},\
		and\ \bibinfo {author} {\bibfnamefont {S.}~\bibnamefont {Gr\"oblacher}},\
	}\bibfield  {title} {\bibinfo {title} {Hybrid integration of silicon photonic
			devices on lithium niobate for optomechanical wavelength conversion},\ }\href
	{https://doi.org/10.1021/acs.nanolett.0c03980} {\bibfield  {journal}
		{\bibinfo  {journal} {Nano Lett.}\ }\textbf {\bibinfo {volume} {21}},\
		\bibinfo {pages} {529} (\bibinfo {year} {2021})}\BibitemShut {NoStop}%
	\bibitem [{\citenamefont {Southworth}\ \emph {et~al.}(2009)\citenamefont
		{Southworth}, \citenamefont {Barton}, \citenamefont {Verbridge},
		\citenamefont {Ilic}, \citenamefont {Fefferman}, \citenamefont {Craighead},\
		and\ \citenamefont {Parpia}}]{Southworth2009}%
	\BibitemOpen
	\bibfield  {author} {\bibinfo {author} {\bibfnamefont {D.~R.}\ \bibnamefont
			{Southworth}}, \bibinfo {author} {\bibfnamefont {R.~A.}\ \bibnamefont
			{Barton}}, \bibinfo {author} {\bibfnamefont {S.~S.}\ \bibnamefont
			{Verbridge}}, \bibinfo {author} {\bibfnamefont {B.}~\bibnamefont {Ilic}},
		\bibinfo {author} {\bibfnamefont {A.~D.}\ \bibnamefont {Fefferman}}, \bibinfo
		{author} {\bibfnamefont {H.~G.}\ \bibnamefont {Craighead}},\ and\ \bibinfo
		{author} {\bibfnamefont {J.~M.}\ \bibnamefont {Parpia}},\ }\bibfield  {title}
	{\bibinfo {title} {Stress and {S}ilicon {N}itride:\ {A} {C}rack in the
			{U}niversal {D}issipation of {G}lasses},\ }\href
	{https://doi.org/10.1103/PhysRevLett.102.225503} {\bibfield  {journal}
		{\bibinfo  {journal} {Phys.\ Rev.\ Lett.}\ }\textbf {\bibinfo {volume}
			{102}},\ \bibinfo {pages} {225503} (\bibinfo {year} {2009})}\BibitemShut
	{NoStop}%
	\bibitem [{\citenamefont {Gr\"oblacher}\ and\ \citenamefont
		{Norte}(2021)}]{Groeblacher2021}%
	\BibitemOpen
	\bibfield  {author} {\bibinfo {author} {\bibfnamefont {S.}~\bibnamefont
			{Gr\"oblacher}}\ and\ \bibinfo {author} {\bibfnamefont {R.}~\bibnamefont
			{Norte}},\ }\href {https://patents.google.com/patent/NL2023917B1/en}
	{\bibinfo {title} {High-selectivity dry release of dielectric structures}}
	(\bibinfo {year} {2021})\BibitemShut {NoStop}%
	\bibitem [{\citenamefont {Almeida}\ and\ \citenamefont
		{Lipson}(2004)}]{Almeida2004}%
	\BibitemOpen
	\bibfield  {author} {\bibinfo {author} {\bibfnamefont {V.~R.}\ \bibnamefont
			{Almeida}}\ and\ \bibinfo {author} {\bibfnamefont {M.}~\bibnamefont
			{Lipson}},\ }\bibfield  {title} {\bibinfo {title} {Optical bistability on a
			silicon chip},\ }\href {https://doi.org/10.1364/OL.29.002387} {\bibfield
		{journal} {\bibinfo  {journal} {Opt.\ Lett.}\ }\textbf {\bibinfo {volume}
			{29}},\ \bibinfo {pages} {2387} (\bibinfo {year} {2004})}\BibitemShut
	{NoStop}%
	\bibitem [{\citenamefont {Camacho}\ \emph {et~al.}(2009)\citenamefont
		{Camacho}, \citenamefont {Chan}, \citenamefont {Eichenfield},\ and\
		\citenamefont {Painter}}]{Camacho2009}%
	\BibitemOpen
	\bibfield  {author} {\bibinfo {author} {\bibfnamefont {R.~M.}\ \bibnamefont
			{Camacho}}, \bibinfo {author} {\bibfnamefont {J.}~\bibnamefont {Chan}},
		\bibinfo {author} {\bibfnamefont {M.}~\bibnamefont {Eichenfield}},\ and\
		\bibinfo {author} {\bibfnamefont {O.}~\bibnamefont {Painter}},\ }\bibfield
	{title} {\bibinfo {title} {Characterization of radiation pressure and thermal
			effects in a nanoscale optomechanical cavity},\ }\href
	{https://doi.org/10.1364/OE.17.015726} {\bibfield  {journal} {\bibinfo
			{journal} {Opt.\ Express}\ }\textbf {\bibinfo {volume} {17}},\ \bibinfo
		{pages} {15726} (\bibinfo {year} {2009})}\BibitemShut {NoStop}%
	\bibitem [{\citenamefont {Johnson}\ \emph {et~al.}(2002)\citenamefont
		{Johnson}, \citenamefont {Ibanescu}, \citenamefont {Skorobogatiy},
		\citenamefont {Weisberg}, \citenamefont {Joannopoulos},\ and\ \citenamefont
		{Fink}}]{Johnson2002}%
	\BibitemOpen
	\bibfield  {author} {\bibinfo {author} {\bibfnamefont {S.~G.}\ \bibnamefont
			{Johnson}}, \bibinfo {author} {\bibfnamefont {M.}~\bibnamefont {Ibanescu}},
		\bibinfo {author} {\bibfnamefont {M.~A.}\ \bibnamefont {Skorobogatiy}},
		\bibinfo {author} {\bibfnamefont {O.}~\bibnamefont {Weisberg}}, \bibinfo
		{author} {\bibfnamefont {J.~D.}\ \bibnamefont {Joannopoulos}},\ and\ \bibinfo
		{author} {\bibfnamefont {Y.}~\bibnamefont {Fink}},\ }\bibfield  {title}
	{\bibinfo {title} {Perturbation theory for {M}axwell’s equations with
			shifting material boundaries},\ }\href
	{https://doi.org/10.1103/PhysRevE.65.066611} {\bibfield  {journal} {\bibinfo
			{journal} {Phys.\ Rev.\ E}\ }\textbf {\bibinfo {volume} {65}},\ \bibinfo
		{pages} {066611} (\bibinfo {year} {2002})}\BibitemShut {NoStop}%
	\bibitem [{\citenamefont {Scheeper}\ \emph {et~al.}(1992)\citenamefont
		{Scheeper}, \citenamefont {Voorthuyzen}, \citenamefont {Olthuis},\ and\
		\citenamefont {Bergveld}}]{Scheeper1992}%
	\BibitemOpen
	\bibfield  {author} {\bibinfo {author} {\bibfnamefont {P.~R.}\ \bibnamefont
			{Scheeper}}, \bibinfo {author} {\bibfnamefont {J.~A.}\ \bibnamefont
			{Voorthuyzen}}, \bibinfo {author} {\bibfnamefont {W.}~\bibnamefont
			{Olthuis}},\ and\ \bibinfo {author} {\bibfnamefont {P.}~\bibnamefont
			{Bergveld}},\ }\bibfield  {title} {\bibinfo {title} {Investigation of
			attractive forces between {PECVD} silicon nitride microstructures and an
			oxidized silicon substrate},\ }\href
	{https://doi.org/10.1016/0924-4247(92)80126-N} {\bibfield  {journal}
		{\bibinfo  {journal} {Sens.\ Actuator A Phys.}\ }\textbf {\bibinfo {volume}
			{30}},\ \bibinfo {pages} {231} (\bibinfo {year} {1992})}\BibitemShut
	{NoStop}%
	\bibitem [{\citenamefont {Bowen}\ and\ \citenamefont
		{Milburn}(2015)}]{Bowen2015}%
	\BibitemOpen
	\bibfield  {author} {\bibinfo {author} {\bibfnamefont {W.~P.}\ \bibnamefont
			{Bowen}}\ and\ \bibinfo {author} {\bibfnamefont {G.~J.}\ \bibnamefont
			{Milburn}},\ }\href {https://doi.org/10.1201/b19379} {\emph {\bibinfo {title}
			{Quantum optomechanics}}}\ (\bibinfo  {publisher} {CRC press},\ \bibinfo
	{year} {2015})\BibitemShut {NoStop}%
	\bibitem [{\citenamefont {Eichenfield}\ \emph {et~al.}(2009)\citenamefont
		{Eichenfield}, \citenamefont {Camacho}, \citenamefont {Chan}, \citenamefont
		{Vahala},\ and\ \citenamefont {Painter}}]{Eichenfield2009a}%
	\BibitemOpen
	\bibfield  {author} {\bibinfo {author} {\bibfnamefont {M.}~\bibnamefont
			{Eichenfield}}, \bibinfo {author} {\bibfnamefont {R.}~\bibnamefont
			{Camacho}}, \bibinfo {author} {\bibfnamefont {J.}~\bibnamefont {Chan}},
		\bibinfo {author} {\bibfnamefont {K.~J.}\ \bibnamefont {Vahala}},\ and\
		\bibinfo {author} {\bibfnamefont {O.}~\bibnamefont {Painter}},\ }\bibfield
	{title} {\bibinfo {title} {A picogram- and nanometre-scale photonic-crystal
			optomechanical cavity},\ }\href {https://doi.org/10.1038/nature08061}
	{\bibfield  {journal} {\bibinfo  {journal} {Nature}\ }\textbf {\bibinfo
			{volume} {459}},\ \bibinfo {pages} {550} (\bibinfo {year}
		{2009})}\BibitemShut {NoStop}%
	\bibitem [{\citenamefont {Krause}\ \emph {et~al.}(2015)\citenamefont {Krause},
		\citenamefont {Blasius},\ and\ \citenamefont {Painter}}]{Krause2015}%
	\BibitemOpen
	\bibfield  {author} {\bibinfo {author} {\bibfnamefont {A.~G.}\ \bibnamefont
			{Krause}}, \bibinfo {author} {\bibfnamefont {T.~D.}\ \bibnamefont
			{Blasius}},\ and\ \bibinfo {author} {\bibfnamefont {O.}~\bibnamefont
			{Painter}},\ }\bibfield  {title} {\bibinfo {title} {Optical read out and
			feedback cooling of a nanostring optomechanical cavity},\ }\href
	{https://arxiv.org/abs/1506.01249} {\bibfield  {journal} {\bibinfo  {journal}
			{arXiv:1506.01249}\ } (\bibinfo {year} {2015})}\BibitemShut {NoStop}%
	\bibitem [{\citenamefont {Wang}\ \emph {et~al.}(2019)\citenamefont {Wang},
		\citenamefont {Nayak},\ and\ \citenamefont {Keloth}}]{Wang2019b}%
	\BibitemOpen
	\bibfield  {author} {\bibinfo {author} {\bibfnamefont {J.}~\bibnamefont
			{Wang}}, \bibinfo {author} {\bibfnamefont {K.~P.}\ \bibnamefont {Nayak}},\
		and\ \bibinfo {author} {\bibfnamefont {J.}~\bibnamefont {Keloth}},\
	}\bibfield  {title} {\bibinfo {title} {Photothermal tuning and stabilization
			of a photonic crystal nanofiber cavity},\ }\href
	{https://doi.org/10.1364/OL.44.003996} {\bibfield  {journal} {\bibinfo
			{journal} {Opt. Lett.}\ }\textbf {\bibinfo {volume} {44}},\ \bibinfo {pages}
		{3996} (\bibinfo {year} {2019})}\BibitemShut {NoStop}%
	\bibitem [{\citenamefont {Jiang}\ \emph {et~al.}(2020)\citenamefont {Jiang},
		\citenamefont {Sarabalis}, \citenamefont {Dahmani}, \citenamefont {Patel},
		\citenamefont {Mayor}, \citenamefont {McKenna}, \citenamefont {{Van Laer}},\
		and\ \citenamefont {Safavi-Naeini}}]{Jiang2020}%
	\BibitemOpen
	\bibfield  {author} {\bibinfo {author} {\bibfnamefont {W.}~\bibnamefont
			{Jiang}}, \bibinfo {author} {\bibfnamefont {C.~J.}\ \bibnamefont
			{Sarabalis}}, \bibinfo {author} {\bibfnamefont {Y.~D.}\ \bibnamefont
			{Dahmani}}, \bibinfo {author} {\bibfnamefont {R.~N.}\ \bibnamefont {Patel}},
		\bibinfo {author} {\bibfnamefont {F.~M.}\ \bibnamefont {Mayor}}, \bibinfo
		{author} {\bibfnamefont {T.~P.}\ \bibnamefont {McKenna}}, \bibinfo {author}
		{\bibfnamefont {R.}~\bibnamefont {{Van Laer}}},\ and\ \bibinfo {author}
		{\bibfnamefont {A.~H.}\ \bibnamefont {Safavi-Naeini}},\ }\bibfield  {title}
	{\bibinfo {title} {{Efficient bidirectional piezo-optomechanical transduction
				between microwave and optical frequency}},\ }\href
	{https://doi.org/10.1038/s41467-020-14863-3} {\bibfield  {journal} {\bibinfo
			{journal} {Nature Commun.}\ }\textbf {\bibinfo {volume} {11}},\ \bibinfo
		{pages} {1166} (\bibinfo {year} {2020})}\BibitemShut {NoStop}%
	\bibitem [{\citenamefont {Saulson}(1990)}]{Saulson1990}%
	\BibitemOpen
	\bibfield  {author} {\bibinfo {author} {\bibfnamefont {P.~R.}\ \bibnamefont
			{Saulson}},\ }\bibfield  {title} {\bibinfo {title} {Thermal noise in
			mechanical experiments},\ }\href {https://doi.org/10.1103/PhysRevD.42.2437}
	{\bibfield  {journal} {\bibinfo  {journal} {Phys.\ Rev.\ D}\ }\textbf
		{\bibinfo {volume} {42}},\ \bibinfo {pages} {2437} (\bibinfo {year}
		{1990})}\BibitemShut {NoStop}%
	\bibitem [{\citenamefont {Imboden}\ and\ \citenamefont
		{Mohanty}(2014)}]{Imboden2014}%
	\BibitemOpen
	\bibfield  {author} {\bibinfo {author} {\bibfnamefont {M.}~\bibnamefont
			{Imboden}}\ and\ \bibinfo {author} {\bibfnamefont {P.}~\bibnamefont
			{Mohanty}},\ }\bibfield  {title} {\bibinfo {title} {Dissipation in
			nanoelectromechanical systems},\ }\href
	{https://doi.org/10.1016/j.physrep.2013.09.003} {\bibfield  {journal}
		{\bibinfo  {journal} {Phys.\ Rep.}\ }\textbf {\bibinfo {volume} {534}},\
		\bibinfo {pages} {89} (\bibinfo {year} {2014})}\BibitemShut {NoStop}%
	\bibitem [{\citenamefont {Fedorov}\ \emph {et~al.}(2018)\citenamefont
		{Fedorov}, \citenamefont {Sudhir}, \citenamefont {Schilling}, \citenamefont
		{Sch\"utz}, \citenamefont {Wilson},\ and\ \citenamefont
		{Kippenberg}}]{Fedorov2018}%
	\BibitemOpen
	\bibfield  {author} {\bibinfo {author} {\bibfnamefont {S.~A.}\ \bibnamefont
			{Fedorov}}, \bibinfo {author} {\bibfnamefont {V.}~\bibnamefont {Sudhir}},
		\bibinfo {author} {\bibfnamefont {R.}~\bibnamefont {Schilling}}, \bibinfo
		{author} {\bibfnamefont {H.}~\bibnamefont {Sch\"utz}}, \bibinfo {author}
		{\bibfnamefont {D.~J.}\ \bibnamefont {Wilson}},\ and\ \bibinfo {author}
		{\bibfnamefont {T.~J.}\ \bibnamefont {Kippenberg}},\ }\bibfield  {title}
	{\bibinfo {title} {Evidence for structural damping in a high-stress silicon
			nitride nanobeam and its implications for quantum optomechanics},\ }\href
	{https://doi.org/10.1016/j.physleta.2017.05.046} {\bibfield  {journal}
		{\bibinfo  {journal} {Phys.\ Lett.\ A}\ }\textbf {\bibinfo {volume} {382}},\
		\bibinfo {pages} {2251} (\bibinfo {year} {2018})}\BibitemShut {NoStop}%
	\bibitem [{\citenamefont {Cartis}\ \emph {et~al.}(2019)\citenamefont {Cartis},
		\citenamefont {Fiala}, \citenamefont {Marteau},\ and\ \citenamefont
		{Roberts}}]{Cartis2019}%
	\BibitemOpen
	\bibfield  {author} {\bibinfo {author} {\bibfnamefont {C.}~\bibnamefont
			{Cartis}}, \bibinfo {author} {\bibfnamefont {J.}~\bibnamefont {Fiala}},
		\bibinfo {author} {\bibfnamefont {B.}~\bibnamefont {Marteau}},\ and\ \bibinfo
		{author} {\bibfnamefont {L.}~\bibnamefont {Roberts}},\ }\bibfield  {title}
	{\bibinfo {title} {Improving the flexibility and robustness of model-based
			derivative-free optimization solvers},\ }\href
	{https://doi.org/10.1145/3338517} {\bibfield  {journal} {\bibinfo  {journal}
			{ACM Trans.\ Math.\ Softw.}\ }\textbf {\bibinfo {volume} {45}},\ \bibinfo
		{pages} {32} (\bibinfo {year} {2019})}\BibitemShut {NoStop}%
\end{thebibliography}

%

\setcounter{figure}{0}
\renewcommand{\thefigure}{S\arabic{figure}}
\setcounter{equation}{0}
\renewcommand{\theequation}{S\arabic{equation}}

\clearpage

\section{Supplementary Information}

\subsection{Assembling procedure}\label{ss:procedure}

\begin{figure*}[!t]
    \centering
    \includegraphics[width=\textwidth]{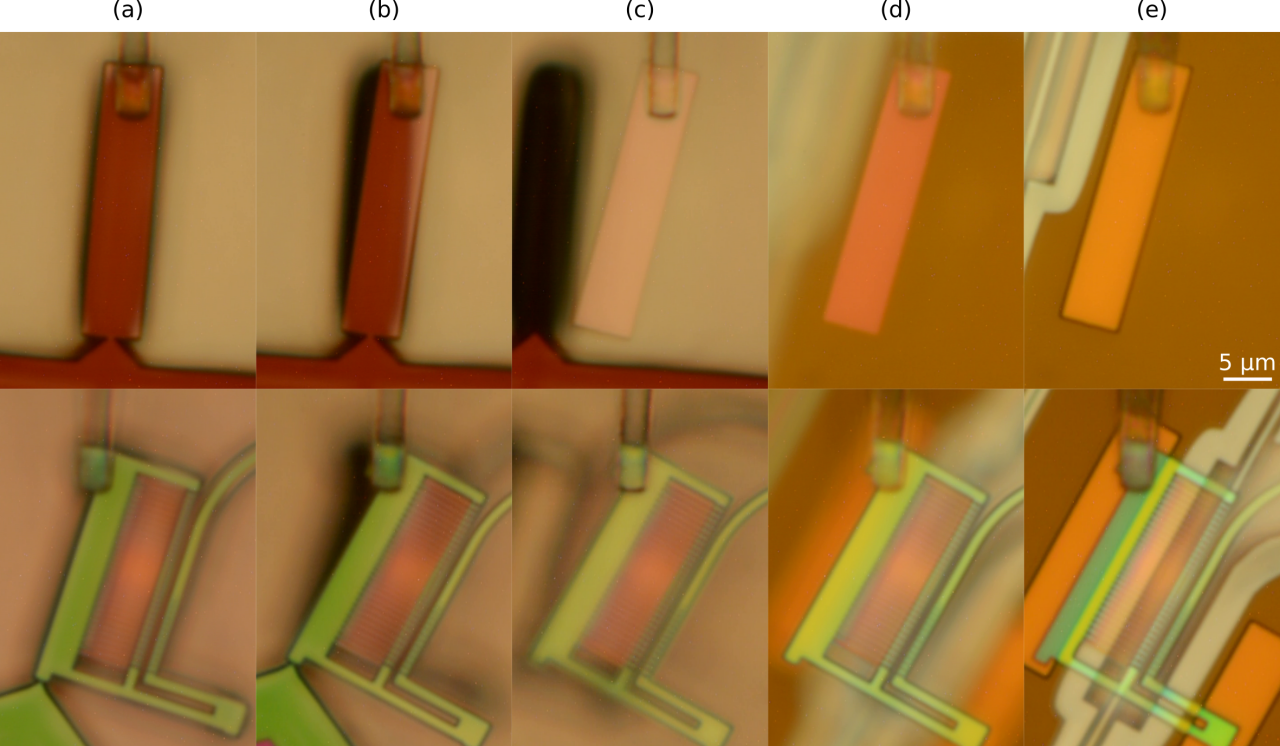}
    \caption{Device assembling procedure for the spacers (top) and the photonic crystal (bottom).}
    \label{fig:assembpics}
\end{figure*}

The optical images of the full assembling procedure of a device are shown in Figure~\ref{fig:assembpics}, with the top rows showing the pick-and-place of the spacer, while the bottom row shows the same transfer process for the photonic crystal. In order to start the assembly, a chip with the spacers, suspended and weakly attached to the substrate, is first placed under the microscope and monitored by a camera. An optical fiber is then used to bend the weak connection between the spacer and the substrate until the connection breaks. The spacer is then positioned above the mechanical chip. Rotational alignment is done beforehand by fitting the edge of the spacer and by comparing it to alignment markers close to the mechanical structure (not shown). The chip is then raised slowly until the spacer touches down. A change in color is visible when this process is completed. These steps are repeated for several spacers. The transfer of the photonic crystal follows the same procedure with the fiber only touching the side of the photonic crystal, avoiding any contamination or damage. Note that for a large photonic crystal structure, not all of it touches down simultaneously. This is noticeable from the color difference for various parts in Figure~\ref{fig:assembpics}(e). In this case, an additional step is performed to push down the other parts of the photonic crystal in order to stick onto the spacers.

\subsection{Reduction of mechanical quality factor}\label{ss:reductionQfactor}

As mentioned in the main text, we typically see a small reduction of the mechanical quality factor after the full structure is assembled. While the exact cause is still not fully understood, we can attribute it to the coupling between the mechanical structure and the photonic crystal, mediated either by charges on the structures, or by van der Waals forces. As the fundamental mechanical mode of the photonic crystal (shown in Figure~\ref{fig:MechSpectMechModes}(c)) has relatively low $Q_\Mech$, this results in an additional loss channel for the mechanical structure.

If we label the displacement of the photonic crystal as $\delta z_\PhC$, and the displacement of the mechanical resonator as $\delta z_\Mech$, we can write the general coupling between the structures as
\begin{equation}
    F_\PhC^c = - F_\Mech^c \approx F_0^c + \alpha_c (\delta z_\PhC - \delta z_\Mech).
\end{equation}
Here, $F_\PhC^c$ is the force exerted from the mechanical resonator on the photonic crystal, and $F_\Mech^c$ is the force on the mechanical resonator. $F_0^c$ is the force at equilibrium, and $\alpha_c$ is the linear coupling strength. Higher order terms to the displacement are neglected for small displacement. As we are only interested in the dynamical behavior, we can neglect $F_0^c$ as it only introduces a constant shift for the equilibrium position.

We further approximate the mechanical resonator and the mechanical motion of the photonic crystal as damped harmonic oscillators and write their equations of motion in the Fourier domain
\begin{equation}
    \begin{gathered}
        \chi_\PhC^{-1} (\omega) \delta z_\PhC (\omega) =  \alpha_c (\delta z_\PhC (\omega) - \delta z_\Mech (\omega)) + F_\PhC (\omega), \\
        \chi_\Mech^{-1} (\omega) \delta z_\Mech (\omega) = - \alpha_c (\delta z_\PhC (\omega) - \delta z_\Mech (\omega)) + F_\Mech (\omega).
    \end{gathered}
\end{equation}
\begin{equation}
    \chi_j^{-1} (\omega) = m_j \left(\Omega_j^2 - \omega^2 - \mathrm{i} \Gamma_j \omega \right)
\end{equation}
is the inverse of the susceptibility of the mechanical resonator ($j=\Mech$) and of the photonic crystal ($j = \PhC$). $F_j$ is the force from other sources, such as the thermal force. In a practical structure, multiple modes present. For the specific geometry of the photonic crystal, the three lowest order modes do not have a large frequency difference and thus our simplified analysis only serve to understand the behavior phenomenologically. We further acknowledge that the dissipation $\Gamma_j$ is also frequency-dependent for internal damping~\cite{Saulson1990,Imboden2014,Fedorov2018}. As we will only be focusing on a small frequency range around $\Omega_\Mech$ we drop this frequency dependency. For a purely internal damping model it could however be different from $\Gamma_\PhC(\omega=\Omega_\PhC)$ by a factor $\Omega_\PhC/\Omega_\Mech$. This factor can be regarded as a constant due to the small $\Gamma_\Mech$. For the whole system, we can then write
\begin{equation}
    \begin{pmatrix}
        \chi_\PhC^{-1} (\omega) - \alpha_c & \alpha_c \\
        \alpha_c & \chi_\Mech^{-1} - \alpha_c
    \end{pmatrix}
    \begin{pmatrix}
        \delta z_\PhC (\omega) \\ \delta z_\Mech (\omega)
    \end{pmatrix} =
    \begin{pmatrix}
        F_\PhC (\omega) \\ F_\Mech (\omega)
    \end{pmatrix}.
\end{equation}

Inverting the matrix on the left-hand side we get
\begin{equation}
    \chi = 
    \begin{pmatrix}
        \left( \chi_\PhC^{-1} + \frac{\alpha_c}{\alpha_c \chi_\Mech - 1} \right)^{-1} &
        \left(\chi_\PhC^{-1} + \chi_\Mech^{-1} - \frac{\chi_\PhC^{-1} \chi_\Mech^{-1}}{\alpha} \right)^{-1} \\
        \left(\chi_\PhC^{-1} + \chi_\Mech^{-1} - \frac{\chi_\PhC^{-1} \chi_\Mech^{-1}}{\alpha} \right)^{-1} &
        \left( \chi_\Mech^{-1} + \frac{\alpha_c}{\alpha_c \chi_\PhC - 1} \right)^{-1}
    \end{pmatrix},
\end{equation}
\begin{equation}
    \begin{pmatrix}
        \delta z_\PhC (\omega) \\ \delta z_\Mech (\omega)
    \end{pmatrix} = \chi (\omega)
    \begin{pmatrix}
        F_\PhC (\omega) \\ F_\Mech (\omega)
    \end{pmatrix}.
\end{equation}
While the two modes are coupled, for our device ($\Omega_\PhC \gg \Omega_\Mech$) we can still approximate them as separate modes around $\Omega_\Mech$
\begin{equation}
    \delta z_\Mech (\omega) \approx \left( \chi_\Mech^{-1} (\omega) + \frac{\alpha_c}{\alpha_c \chi_\PhC (\omega) - 1} \right)^{-1} F_\Mech(\omega).
\end{equation}
as the off-diagonal terms are much smaller in magnitude. Then, for the mechanical resonator, the inverse of effective susceptibility becomes
\begin{equation}
    1 / \chi_\Mech^{\mathrm{eff}} (\omega) = \chi_\Mech^{-1} (\omega) + \frac{\alpha_c}{\alpha_c \chi_\PhC (\omega) - 1} .
\end{equation}
We now look at the imaginary part (the dissipation~\cite{Aspelmeyer2014}) and evaluate around the frequency of the mechanical resonator
\begin{equation}
    \begin{aligned}
        & \mathrm{Im} \frac{1} {\chi_\Mech^{\mathrm{eff}} (\omega)} \approx \\
        & - m_\Mech \left(\Gamma_\Mech + \frac{\alpha_c^2 \Gamma_\PhC / m_\Mech m_\PhC }{\left(\alpha_c/m_\PhC - (\Omega_\PhC^2 - \Omega_\Mech^2)\right)^2 + \Gamma_\PhC^2 \Omega_\Mech^2} \right) \omega.
    \end{aligned}
\end{equation}
We can clearly see that the dissipation is increased by
\begin{equation}
    \begin{aligned}
        \delta \Gamma_\Mech &= \frac{\alpha_c^2 \Gamma_\PhC / m_\Mech m_\PhC }{\left(\alpha_c/m_\PhC - (\Omega_\PhC^2 - \Omega_\Mech^2)\right)^2 + \Gamma_\PhC^2 \Omega_\Mech^2} \\
        & \approx
        \frac{\alpha_c^2}{m_1 m_2} \frac{\Gamma_\PhC}{\Omega_\PhC^4}.
    \end{aligned}
\end{equation}

Our model shows that if there is any coupling between the mechanical resonator and the mechanics of the photonic crystal, the quality factor of the mechanical resonator is always reduced. The last approximation is appropriate for our devices since $\Omega_\PhC \approx 8 \Omega_\Mech$ for the fundamental mechanical mode of the photonic crystal. Though the equation shows a $\Omega_\PhC^{-4}$ dependence, we would like to emphasis that $\Gamma_\PhC$ also depends on $\Omega_\PhC$, in particular for structural damping, ($\Gamma_\PhC \sim \Omega_\PhC^2$~\cite{Saulson1990,Imboden2014,Fedorov2018}). Nevertheless, increasing the mechanical frequency of the photonic crystal reduces the dissipation degradation, and is one of the main reasons why the tethers are used. In order to verify this, we have also designed and assembled photonic crystals without the clamping tethers with $\Omega_\PhC \approx 3 \Omega_\Mech$, and see a systematically larger degradation in the mechanical quality factor.

\subsection{Optical bistability}

Optical bistability can either be caused by photon absorption of the material that generates heat, or the static optomechanical effect~\cite{Almeida2004,Aspelmeyer2014}. Since both similarly affect our measurements, we do not need to distinguish them. Furthermore, the extra clamping tethers we introduce can compensate for both as they increase the thermal conductivity and the rigidity of the photonic crystal structure.

In the presence of linear absorption or the static optomechanical effect, the optical resonance frequency shifts and is proportional to the cavity photon number with a coefficient $\beta$,
\begin{equation}
    \delta \omega_\cav = \omega_\cav - \omega_\cav^{(0)} = \beta n_\cav(\Delta),
\end{equation}
where $\omega_\cav^{(0)}$ is the original cavity resonance frequency. In a typical situation, $\beta < 0$. The actual detuning, is then a function of the cavity photon number
\begin{equation}\label{eq:detun}
    \Delta (n_\cav) = \Delta_0 - \beta n_\cav(\Delta).
\end{equation}
$\Delta_0$ is the detuning with vanishing input power and is a fixed value for a given laser wavelength. Let the cavity photon number at resonance be $N_\cav$, which is given by 
\begin{equation}\label{eq:ncav}
    n_\cav(\Delta) = \frac{(\kappa/2)^2}{(\kappa/2)^2 + \Delta^2} N_\cav,
\end{equation}
for a fixed input power. Equation~\eqref{eq:detun} and \eqref{eq:ncav} define the behavior of the system, and we use them to fit the traces in Figure~\ref{fig:ThermalTun}(a) in the main text. Combining these two equations, we get an equation for $\Delta$,
\begin{equation}\label{eq:bistab_detun_eq}
    \Delta^3 - \Delta_0 \Delta^2 + \left(\frac{\kappa}{2}\right)^2 \Delta + \left(\frac{\kappa}{2}\right)^2 \left( \beta N_\cav - \Delta_0 \right) = 0
\end{equation}
Optical bistability occurs when there are more than one possible $\Delta \in R$ for a fixed laser frequency, namely a fixed $\Delta_0$. Consider a situation when the laser is on resonance, $\Delta = 0$. A small perturbation, either slightly reducing the photon number or $\Delta$ becoming slightly smaller than 0, the cavity photon number is reduced and hence the optical resonance frequency drifts to the high frequency side. This then reduces the cavity photon number and makes $\Delta$ more negative, forming a positive feedback loop. The system would then abruptly transit to a state with another $\Delta$, if it exists, where the optical cavity resonance is very far away and the cavity photon number is small. The system would not come back to the original state with any further perturbations. To avoid this instability, it has been proposed to always work at an optical wavelength where multiple values for $\Delta$ are not possible~\cite{Wang2019b}. However, as this is in the blue-detuned regime, it can lead to coherent driving and heating of the mechanical resonator, and the system becomes unstable again when a high $Q_\Mech$ mechanical resonator is involved. On the other hand, if there is only one feasible $\Delta$, the state of the system is still close to the original state. This process is continuous, and a small change of the environment can still bring the system back. Thus, the transition between multiple solutions and a single solution in equation~\eqref{eq:bistab_detun_eq} marks the boundary between a stable and an unstable system.

Let us now define 
\begin{equation}
    y (\Delta) = \Delta^3 - \Delta_0 \Delta^2 + \left(\frac{\kappa}{2}\right)^2 \Delta + \left(\frac{\kappa}{2}\right)^2 \left( \beta N_\cav - \Delta_0 \right).
\end{equation}
For equation~\eqref{eq:bistab_detun_eq}, having multiple solutions implies that $y$ is not a monotonically increasing function and $y^\prime (\Delta)$ should have 2 real solutions. Thus, $\Delta_0^2 - \frac{3}{4} \kappa^2 > 0$ and the local minimum and maximum happens when
\begin{equation}
    \Delta = \Delta_\pm = \frac{1}{3} \Delta_0 \pm \frac{1}{6} \sqrt{4 \Delta_0^2 - 3 \kappa^2}.
\end{equation}
Then, the bistability occurs if and only if there exists a real $\epsilon_0$ such that $y(\Delta_+) < 0$ and $y(\Delta_-) > 0$. Note that both $y(\Delta_+)$ and $y(\Delta_-)$ are monotonically decreasing as $\Delta_0$ increases in their ranges. The bistability bound is then given by
\begin{equation}
    N_\cav > \left| \frac{4 \sqrt{3} \kappa}{9 \beta} \right|.
\end{equation}
The non-stable regime is colored in gray in Figure~\ref{fig:ThermalTun}(b), with $\beta$ obtained by the linear fit of the change of optical resonance frequency. We note that the measurement in the bistability regime might not be as reliable. We first perform a linear fit to all the data points in Figure~\ref{fig:ThermalTun}b, extracting a bistability bound of $N_\cav \approx 3000$. We then perform another fitting with only the data $N_\cav < 3000$ and do not see a substantial difference between the two fits. The fitting result and the corresponding bistability bound is shown in Figure~\ref{fig:ThermalTun}(b).

\begin{figure}[!t]
    \centering
    \includegraphics[width=1.\columnwidth]{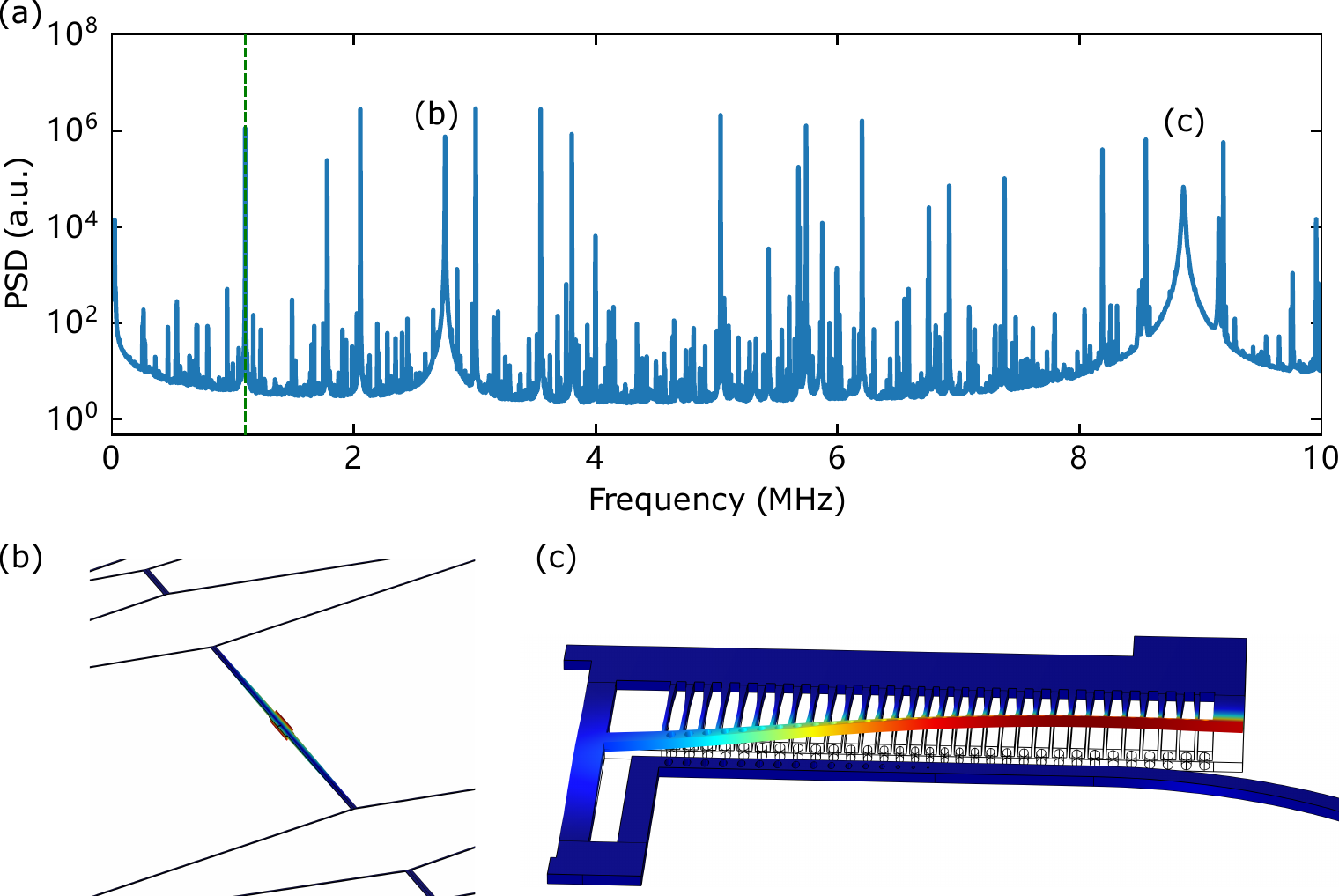}
    \caption{(a) Mechanical spectrum over a large frequency range. The fundamental mode (marked in green) has a high mechanical quality factor and large optomechanical coupling to the optical cavity. (b) and (c) show a torsional mode of the mechanical structure and the fundamental mechanical mode of the photonic crystal, respectively.}
    \label{fig:MechSpectMechModes}
\end{figure}

\subsection{Device characterization}

\begin{figure}[!t]
    \centering
    \includegraphics[width=1.\columnwidth]{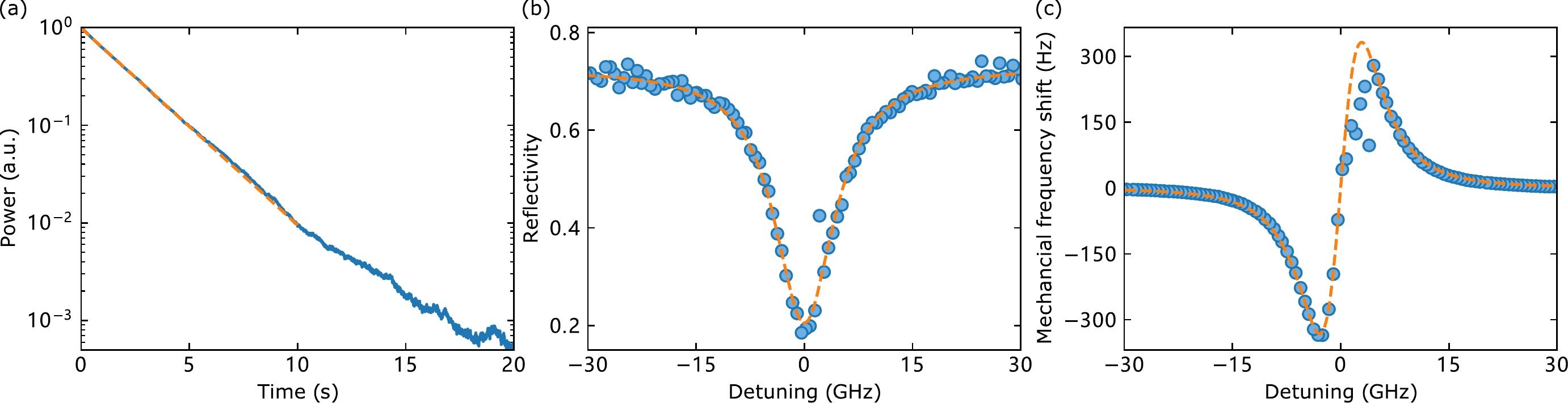}
    \caption{Characterization of one of our devices. (a) A ringdown measurement shows a quality factor of $Q_\Mech = 1.49 \times 10^7$ ($\Omega_\Mech/2\pi = 1.10$~MHz). (b) Unnormalized cavity reflection around the optical resonance (1554.6~nm). The fitting yields a linewidth $\kappa/2\pi= 10.1$~GHz. (c) Optical spring measurement by detuning the input laser around the cavity resonance frequency. On the blue detuning side ($\omega_\mathrm{L} > \omega_\mathrm{cav}$), the mechanical frequency is not stable due to optical heating, resulting in a slightly non-ideal fit to the data.}
    \label{fig:singledevice_char}
\end{figure}

Exemplary data optained from the device characterization is shown in Figure~\ref{fig:singledevice_char}, including a ringdown measurement for the mechanical resonator after assembling the full device, an optical cavity reflection measurement and an optical spring measurement. From the fits we obtain a quality factor of $Q_\Mech = 1.49 \times 10^7$ and $g_0/2\pi = 257.4 \pm 4.9$~kHz.

\subsection{Mechanical resonator design}

The mechanical resonator is formed by a long string and a fractal structure. On each level of the fractal, a set of parameters, including the tether widths, lengths, the size of the diamond-shape connections, are assigned. This forms a large parameter space and optimization is performed using the Python package Py-BOBYQA~\cite{Cartis2019}. It implements a derivative-free optimization algorithm to minimize the objective function (mechanical quality factor) within a bounded parameter space. It approximates a small region in the parameter space by a quadratic function, and then seeks the next point by interpolation. Due to the expensive simulation of the mechanics and the large parameter space, the optimization is terminated when the quality factor is ``good enough''. For the mechanical structure, it still has in-plane and torsional modes. Both can have low mechanical quality factor, and their thermal motion might present as excess classical noise in sensitive measurements. We thus also want the frequency of the fundamental in-plane mode and the torsional mode to be as far away in frequency as possible. This is however, not included in the optimization part and is performed manually by enlarging the center diamond.

\end{document}